\documentclass[twocolumn,english,aps,prb,showpacs,byrevtex,amsmath,amssymb,superscriptaddress]{revtex4-1}
\usepackage{makecell}
\usepackage[colorlinks,allcolors=black,pdftex]{hyperref}
\usepackage{dcolumn}
\usepackage{bm}
\usepackage{amsfonts}
\usepackage{latexsym}
\usepackage{import}
\usepackage{color}
\usepackage{hyperref}
\usepackage[normalem]{ulem}
\usepackage{fancyhdr}
\usepackage{amsbsy}
\usepackage{amsmath}
\usepackage{epsfig}
\usepackage{amssymb}
\usepackage{pifont}
\usepackage{textcomp}
\usepackage{gensymb}
\usepackage{pifont}
\usepackage{enumerate}
\usepackage[T1]{fontenc}
\usepackage{graphicx}
\usepackage[title]{appendix}
\usepackage{bm}
\usepackage{multirow}
\usepackage{caption}
\usepackage{subcaption}
\usepackage{titlesec}
\titlespacing\subsection{10pt}{10pt plus 4pt minus 2pt}{10pt plus 2pt minus 2pt}

\usepackage{tikz,xcolor,hyperref}
\definecolor{lime}{HTML}{A6CE39}
\DeclareRobustCommand{\orcidicon}{%
	\begin{tikzpicture}
	\draw[lime, fill=lime] (0,0)
	circle [radius=0.16]
	node[white] {{\fontfamily{qag}\selectfont \tiny ID}};
	\draw[white, fill=white] (-0.0625,0.095)
	circle [radius=0.007];
	\end{tikzpicture}
	\hspace{-2mm}
}

\foreach \x in {A, ..., Z}{%
	\expandafter\xdef\csname orcid\x\endcsname{\noexpand\href{https://orcid.org/\csname orcidauthor\x\endcsname}{\noexpand\orcidicon}}
}

\usepackage{hyperref}
\makeatletter
\newcommand{\printfnsymbol}[1]{%
  \textsuperscript{\@fnsymbol{#1}}%
}
\makeatother

\begin{document}

\begin{abstract}
Thermoelectricity is a next-generation solution for efficient waste heat management. Although various thermoelectric materials exist, there is still a lot of scope for advancement, especially in room temperature applications. Recently, two-dimensional (2D) materials, including MXenes, showed promise as thermoelectric materials. On the other hand, MXenes generally exhibit metallic behavior that can hinder thermoelectric performance.  Nevertheless, the variety of MXene's chemical composition and surface functionalization facilitate the research path based on energy band engineering, carrier concentration, and mobility. Multiple strategies to enhance the thermoelectric properties of layered MXenes materials, including structural modifications, defects, band gap engineering, etc. have been comprehensively demonstrated. In addition, advanced structural engineering such as nanostructuring MXenes with materials of different dimensions, creating van der Waals heterostructures, alloying, and utilizing MXenes as nanoinclusions or nanocomposites is presented. The thermoelectric efficiency of MXenes over the landscape of other 2D and conventional thermoelectric materials has been systematically compared. Meanwhile, a future approach has been proposed to enhance the thermoelectric properties of novel members of the flatland, MBenes exhibiting an incredible diversity of structures and crystal symmetries. Finally, potential applications in thermoelectrics and future prospects of MXenes are discussed. This article provides a timely and unique review of MXenes’ advantages and limitations that have never been so well understood and established. This creates a comfort zone for rational tailoring of their structure-property-performance relationship, which is crucial concerning the thermoelectric performance, widely covered in this review. 
\end{abstract}

\title{Recent progress in thermoelectric MXene-based structures versus other 2D materials}

\author{Subrahmanyam Bandaru
\orcidA}\thanks{These authors contributed equally.}
\affiliation{Faculty of Physics, University of Warsaw, Pasteura 5, PL-02093 Warsaw, Poland}
\author{Agnieszka M. Jastrzębska}
\affiliation{Warsaw University of Technology, Faculty of Materials Science and Engineering, 02-507 Warsaw, Wołoska 141, Poland}
\author{Magdalena Birowska}\thanks{Corresponding author}
\affiliation{Faculty of Physics, University of Warsaw, Pasteura 5, PL-02093 Warsaw, Poland}
\maketitle{}

\section{Introduction}

Energy has a crucial role in the development of civilization. It is produced from unsustainable nonrenewable fossil fuels such as coal, petroleum, and gas \cite{AFSHAR20125639, KALKAN20126352, XI2007923}. While recent global energy consumption is retarded due to the Covid-19 pandemic, further global crises associated with Russia’s invasion of Ukraine exposed the bottlenecks of current energy management. The energy cycle starts at power plants and goes through factories into final consumables. Unfortunately, this process has an open end where massive heat is wasted \cite{MARTINGONZALEZ2013288, SHU2013385, WANG2011}. Therefore, it is essential to explore novel renewable energy solutions that could contribute to closing the energy cycle. This can be facilitated by thermoelectric (TE) materials having the ability to recover the released energy.

\begin{figure}[h]
    \centering
\includegraphics[width=0.45\textwidth]{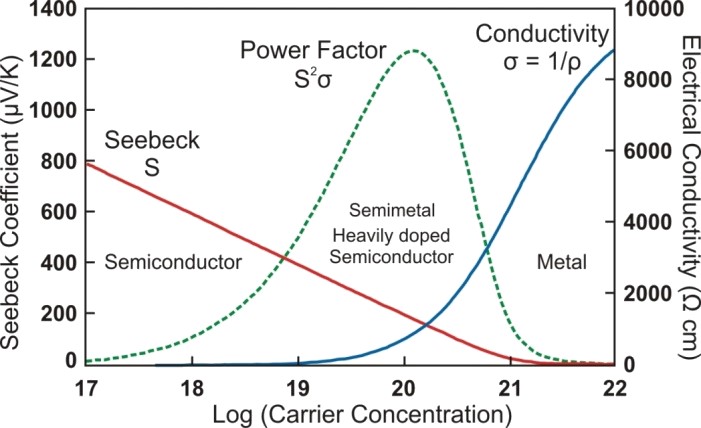}
    \caption{{Variation of transport parameters with carrier concentration}\cite{_Ioffe_1957}}
    \label{ZT_of_materials}
\end{figure}

The TE materials bring the energy back into the cycle due to being capable of converting heat into electricity. This phenomenon is called heat-to-electricity conversion and is associated with Seebeck, Peltier, and Thomson effects. Seebeck effect is a phenomenon in which the temperature difference between two dissimilar conductors leads to the production of the voltage difference between them \cite{Seebeck, Seebeck_rowe}. While an inverse phenomenon to the Seebeck effect, that is the heating effect or cooling effect which can be obtained by the variation of electric flux is known as the Peltier effect \cite{Peltier_effect}. The inter-dependency of the Seebeck and Peltier effects has been applied by the third effect known as the Thomson effect \cite{Thomson_effect}. In the latter effect, the heat is absorbed or released when current flows in a conductor with a temperature gradient. Using these TE effects, for the generation of electricity or refrigeration, one requires thermocouples. A thermocouple is an electric device made of two dissimilar conductors forming an electrical junction which are at different temperatures. A TE device for larger power systems can be constructed from a TE module built from a large number of thermocouples. Notably, TE materials are the heart of their corresponding TE devices.
Since the performance of TE devices is determined by applied TE material, it can only be improved by its engineering. Figure \ref{ZT_of_materials} displays the crucial characteristics required for a good TE material, based on Seebeck coefficient (S), power factor ($S^2\sigma$), electrical conductivity ($\sigma$) in relation to the carrier concentration \cite{_Ioffe_1957}. The engineering aims at increasing the Seebeck coefficient (S) and the electrical conductivity ($\sigma$) while simultaneously lowering its thermal conductivity $\kappa$ (=$\kappa_e + \kappa_l$), where $\kappa_e$ and $\kappa_l$ are electronic thermal conductivity and lattice thermal conductivity, respectively. Due to the complex interrelation among these three parameters, it is almost impossible to optimize them independently. 
Altogether, the engineering of these entities can enhance the efficiency of TE material in generating power or cooling, which can be described by a dimensionless quantity named a figure of merit ZT = ($S^2\sigma/\kappa$)T. In other words, the ideal TE material should have a high ZT value based on a high power factor with high electrical conductivity and low thermal conductivity.

The computational research on TE materials plays a key role in progressing the discovery and expansion of high performance TE materials. The theoretical investigation of TE properties begins with the first-principles calculations of the ground state and electronic band structure using density functional theory (DFT) \cite{David_singh_computational_progress}. This methodology combines both all-electron simulations, using the WIEN2K software \cite{WIEN2K}, as well as other approaches including pseudopotentials codes like VASP \cite{VASP} and Quantum ESPRESSO \cite{QUANTUMESPRESSO}. An excellent and efficient way to analyze the TE properties of materials is Boltzmann’s transport equation (BTE) within the constant relaxation time ($\tau$) approximation using the results from electronic structure calculations. This approach can be applied to obtain the transport quantities, particularly TE properties, using the outputs of electronic structure calculations from DFT. Seebeck coefficient can be computed within this approximation without any dependence on $\tau$, but the electrical conductivity and thus the power factor are computed concerning $\tau$. The constant relaxation time approximation enables the determination of doping (Fermi level) and temperature dependent, S($\mu$, T) entirely from the band structure, eliminating the need for any reliance on experimental data fitting. Employing this method, many cases have revealed the critical impact of nontrivial band structure characteristics on the Seebeck coefficient's evaluation. Based on this approximation, the Seebeck coefficient concerning Fermi-Dirac distribution ($f$), group velocity ($v$), and density of states ($\rho$) can be computed as follows:

\begin{equation}
  S(\mu, T)=\frac{e}{\ohm T \sigma}\int_{}{}d\varepsilon \tau_{el}(\varepsilon,\mu, T) v^2(\varepsilon) \rho(\varepsilon) (\varepsilon - \mu)[-\partial_\varepsilon f_\mu (\varepsilon)] 
\end{equation}

where e, $\ohm$, $\varepsilon$, $\mu$, $\tau_{el}$ are the electron charge, volume of the unit cell, electron energy, chemical potential, and the electronic relaxation time, respectively (for more details see review paper Ref. \cite{David_singh_computational_progress} and references therein). The commonly used computational tool in the TE community for the analysis of electronic transport properties is the BoltzTrap code \cite{BoltzTraP}. Meanwhile, unlike the Seebeck coefficient, the electrical conductivity and the electronic thermal conductivity strongly depend on $\tau_{el}$, and these quantities remain elusive solely through analysis of band structure. The electrical conductivity and the electronic conductivity can be evaluated with the following equations:

\begin{equation}\label{sigma}
\sigma(\mu, T)=\frac{e^2}{\ohm}\int_{}{}d\varepsilon \tau_{el}(\varepsilon,\mu, T) v^2(\varepsilon) \rho(\varepsilon) [-\partial_\varepsilon f_\mu (\varepsilon)]
\end{equation}

\begin{equation}\label{kappa_e}
\begin{split}
\kappa_{el}(\mu, T)=&
 -\sigma T S^2 +\\ &\frac{e}{\ohm T}\int_{}{}d\varepsilon \tau_{el}(\varepsilon,\mu, T) v^2(\varepsilon) \rho(\varepsilon) (\varepsilon - \mu)^2 [-\partial_\varepsilon f_\mu (\varepsilon)] 
 \end{split}
 \end{equation}

Finally, the intrinsic lattice thermal conductivity can be estimated from two distinct approaches,  namely, Boltzmann theory in momentum space or Green-Kubo molecular dynamics in real space. The Boltzmann theory formalism relies on the framework of a weakly interacting gas of phonons and the lattice thermal conductivity can be calculated using the Peierls-Boltzmann transport equation as follows:
 
\begin{equation}\label{kappa_ph}
    \kappa_{ph}(T)=
 \frac{1}{k_B\ohm T^2}\sum_{q}^{BZ}\tau_{ph}(q, T) v_q^2(\hbar w_q)^2 n(w_q)[n(w_q)+1]
\end{equation}

where BZ is the Brillouin zone, $\hbar$ is the Planck's constant, $\tau_{ph}$ is the phonon lifetime, $v_q$ is the phonon group velocity, $w$ is the phonon frequency, and $n$ is the Bose-Einstein distribution. The harmonic and anharmonic force constants required to obtain the lattice thermal conductivity can be calculated using the finite displacements method \cite{DFPT}. Currently, the most popular software tools to compute the lattice thermal conductivity are ShengBTE \cite{ShengBTE} for pristine compounds and almaBTE \cite{almaBTE} for defective structures.

Apart from the computational approaches, the effectiveness of the TE materials is also being investigated by various experimental procedures \cite{TEproperty_measurement_WEI20182183}. The measurement of $\sigma$ and $S$ is most frequently performed using the commercially available instruments, Ulvac-Riko ZEM and Linseis LSR systems. A specimen is placed under a temperature gradient and the temperature difference ($T_1 - T_2$) is measured by thermocouples. $S$ is then estimated by the measurement of the thermal electromotive force ($dE$) generated due to the temperature difference between the upper and lower parts of the sample.

\begin{equation}
  S=\frac{dE}{T_1 - T_2} 
\end{equation}

The electrical resistivity is measured by the standard dc four-terminal method. When a constant current (I) is applied to both ends of the sample, a potential difference $dV$ is measured using thermocouples. The resistivity is measured by the following equation: 

\begin{equation}
  \rho=\frac{dV}{I}.\frac{A}{l} 
\end{equation}

where $l$ is the length of the sample and $A$ is surface area of the sample. Meanwhile, $\kappa$ can be directly measured with the help of apparatus like a physical property measurement system (PPMS) below room temperature. For higher temperatures, $\kappa$ can be determined by the material's thermal diffusivity ($a$), specific heat capacity ($C_p$), and density ($d$) as $\kappa = aC_pd$. The thermal diffusivity of the samples is usually measured by Laser Flash Analysis (LFA). The $C_p$ is more frequently evaluated by the differential scanning calorimetry (DSC) and PPMS methods.

Besides, the evaluation of TE properties from the computational and experimental methods, it is essential to exploit the TE materials for different applications. The ability to convert energy and susceptibility to engineering places TE materials in the spotlight of considerable attention. Consequently, TE devices became a major research field in the middle of the 20th century \cite{DeWitt1939, Shaver1996}. Considerable progress has been made in the area of high-temperature TE materials for application in power plants, automobiles, and industrial technology. The applications of TE materials are advanced to various fields by the aid of TE generators, TE refrigerators, electronic device coolers, automobile system coolers, and many other ways \cite{TE_applications_1}. The TE generators are adopted in power plant industries for waste heat recovery, automotive energy harvesting, aerospace and aviation, solar energy generation, etc \cite{TE_applications_2, TEG_2, TEG_3}. The TE refrigeration using the Peltier effect is utilized in manufacturing economical TE refrigerators, portable solar TE refrigeration systems for outdoor use, and using TE freezers for medical benefits \cite{TER_1, TER_2, TE_coolers}. The TE materials are also used for various medical applications like wearable sensors, brain coolers, skin coolers, implantable TE generators, etc. \cite{TE_medical_1, TE_medical_2}. However, there is no such progress in TE materials for room-temperature applications.

Conventional TE materials such as lead chalcogenides (PbX, where X = S, Se, Te) \cite{PbTe_WU20191276, BiTe_PbTe_hochbaum_enhanced_2008}, bismuth telluride (Bi$_2$Te$_3$) \cite{Bi0.5Sb1.5Te3_2015_Kim} or silicon germanium alloys \cite{Si-Ge_1_TE, Si-Ge_2_TE} possess ZT values close to 1 at room temperature, which is not enough to reach the high efficiency. On the other hand, rare elements such as Te are expensive and toxic, which might prevent their use in large-scale production. Recently, a new strategy has been implemented to design Bi$_2$Te$_3$ thin films with outstanding TE performance and  high flexibility \cite{Bi2Te3_zheng_harvesting_2023}. SnSe polycrystals synthesized by solvothermal synthesis improved their ZT values and simultaneously enhanced their mechanical strength \cite{Solvothermal_ZhiGang}.

Accordingly, many compositions were investigated for TE devices. In particular, the TE effect was found for clatharates, complex chalcogenides \cite{PbTe_WU20191276, chalcogenides, chalcogenides_2}, skutterides \cite{skutterides_1_TE, skutterides_2_TE}, half-Heuslers \cite{Heusler_Graf_2011}, Zintl-phase compounds \cite{Zintl_thermoelectric}, quasicrystals \cite{quasicrystals}, silicides \cite{Silicides} and various oxides. However, their heat-to-electrical conversion efficiency is still too low regarding the current needs. This drives the evolution of TE materials into more advanced and sophisticated compositions.

In this scene, 2D materials are securing their place and stand at the forefront due to their atomically thin nature. Typically, the 2D materials are referred to as crystals composed of single- or few-layer atoms, featuring strong covalent bonds within the plane and relatively weaker van der Waals (vdW) bonds in the out-of-plane direction. From the perspective of materials engineering, 2D materials can also be described as materials with lateral dimensions on the order of micrometers, surpassing the transverse dimensions that commonly range about nanometers. The first observation of changes in physical and chemical properties due to the quantum size effect came along with the exfoliation of graphene from graphite in 2004. This allured the field of nanotechnology towards novel possibilities and advancements \cite{Novoselov_2004, CastroNeto_2009}. This exceptional change in the dimensionality of the material had a huge impact on intrinsic physical properties such as electrical, mechanical, and thermal transport \cite{Graphene_1, Graphene_2}. In addition, other unique properties of 2D materials including large surface area, high versatility, low density, and peculiar regulation over morphology and compositions enable them to be utilized in energy storage applications. 
 
This ignited a spark in the scientific community to hunt for new 2D materials for TE applications. The promising results have been reported for the layered SnSe \cite{Zhao2014}, and experimentally synthesized Sb$_2$Si$_2$Te$_6$ \cite{LUO2020159} which exhibited intrinsically low lattice thermal conductivity and relatively high Seebeck coefficient. Over twenty years, the family of 2D layered structures evolved quickly leading to the successful synthesis of hexagonal boron nitride (h-BN), black phosphorous (BP), silicene, transition metal dichalcogenides (TMDs), transition metal phosphorus trichalcogenides (MPX$_3$) \cite{2DFePs3, MPX3, PhysRevResearch.4.023256}, and many others. About ten years ago, a new family of 2D materials appeared such as early transition metal carbides and nitrides (MXenes) and very recently transition metal borides (MBenes).

In this review article, we discuss the TE performance of MXene-type and other 2D materials. In the following sections, we discuss the advanced approaches such as band engineering including surface functionalization, nanostructuring particular preparation of crystal forms to ensure good TE properties through the utilization of MXenes crystal structures and beyond, then we address briefly various types of 2D materials, and their TE properties.  Further, a minor contribution of MBenes to the field of thermoelectricity to date is reviewed. There is a lot of scope to design MBenes as promising TE materials with the inspiration of MXenes. In one of the last sections, we provide a deep insight into tailoring MBenes into efficient TE materials. Finally, we discuss  insights and challenges
that have to be considered to overcome the limitations in the thermoelectric research field.

\section{Thermoelectricity in MXenes}
\begin{figure*}
    \centering  \includegraphics[width=1.0\textwidth]{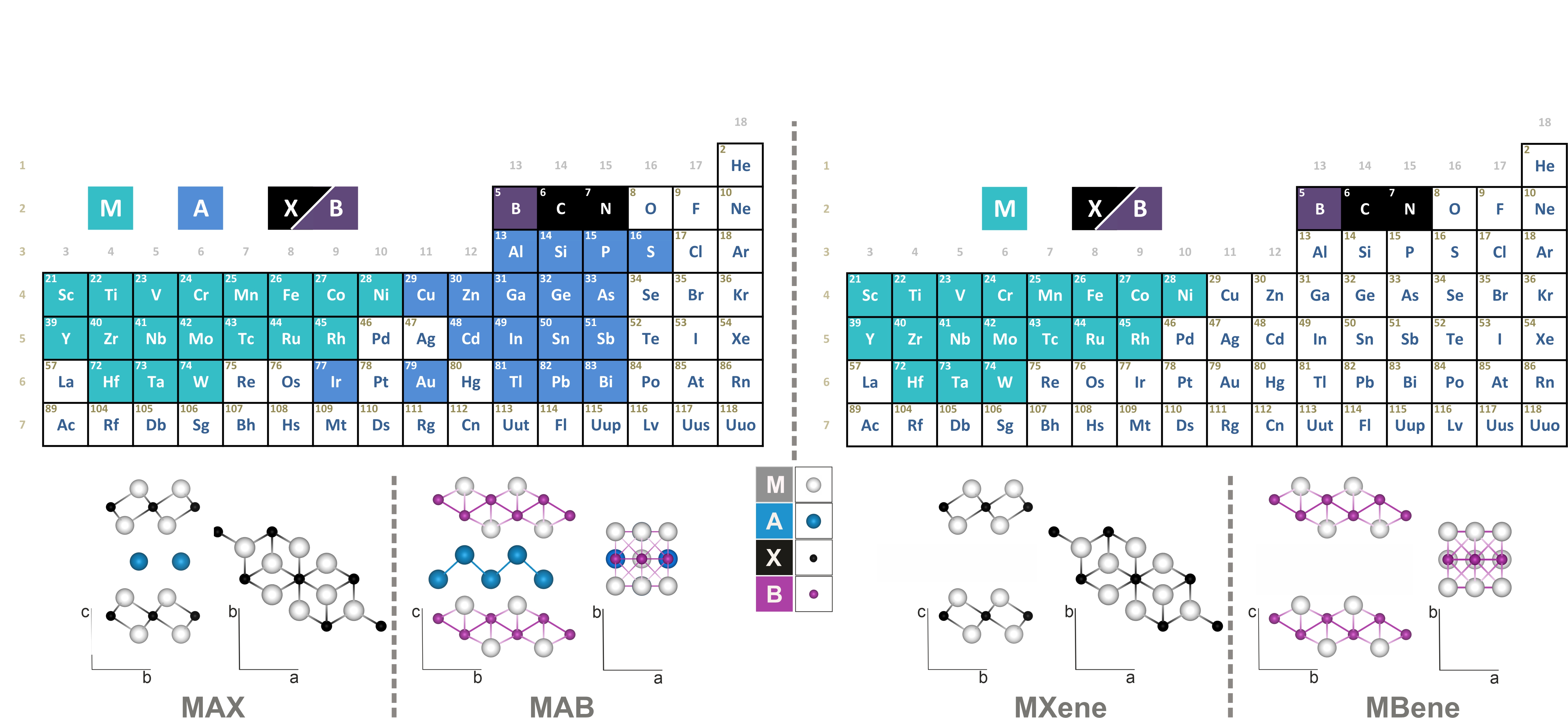}
    \caption{Periodic table of elements illustrating the MAX/MAB phases and their' corresponding MXenes/MBenes structures. The particular crystal structures, hexagonal (MAX/MXene) and orthorhombic (MAB/MBene) are presented at the bottom of the figure.}
    \label{PeriodicTable}
\end{figure*}

The 2D transitional metal carbides and/or nitrides are termed as MXenes\cite{anasori_MXenes_2017}. Since the discovery of Ti$_3$C$_2$ in the year 2011 \cite{https:Ti3C2_Naguib_2011}, the group of MXenes has been developed in several ways. MXenes are formed by exfoliating 'A' from the MAX phases. The MAX phases are a class of layered materials exhibiting metallic properties with a chemical composition of M$_{n+1}$A$X_n$, (n = 1-3), M stands for early transition metal (Sc, Zr, Hf, Mo, Ta, Ti, Hf, V, Nb, etc), A denotes mostly group 13 and 14 elements of the periodic table and X is carbon and /or nitrogen \cite{Naguib_functionalization_2014, MXene_1, MXene_2, MXene_3, MXene_4}. The selective etching of A layers to the formation of MXenes is made possible by the M-A bonds that are chemically more active than the stronger M-X bonds. The general formula of MXenes is M$_{n+1}$X$_n$T$_x$ (n = 1-3), where M stands for early transition metal (Sc, Zr, Hf, Mo, Ta, Ti, Hf, V, Nb, etc), X is carbon and /or nitrogen and T$_x$ stands for the surface terminations (like OH, O or F) \cite{Naguib_functionalization_2014}. Figure \ref{PeriodicTable} shows different types of elements from the periodic table that highlights the formation of MAX phases and MXene compounds.

In addition, the MXenes endure high commitment to renewable energy applications due to their remarkably high electric conductivity, excellent structural and chemical stabilities, high hydrophility and splendid surface chemistry \cite{MXene_applications_1, MXene_applications_2, MXene_applications_3, MXene_applications_4, MXene_applications_5, MXene_applications_6}. The uniqueness of MXenes made them suitable for wide range of applications such as energy storage \cite{MXenes_app_energy_1, MXenes_app_energy_2, MXenes_app_energy_3, MXenes_app_energy_4}, sensing \cite{MXenes_app_sensing_1, MXenes_app_sensing_2}, photodetectors \cite{MXenes_photodetectors}, water purification \cite{MXenes_waterpure_1, MXenes_waterpure_2, MXenes_waterpure_3}, electromagnetic interference shielding \cite{MXenes_EM_1, MXenes_EM_2, MXenes_EM_3}, biomedical applications \cite{MXenes_Bio_1, MXenes_Bio_2}, electronics and optoelectronics applications \cite{MXenes_elec_opto_1, MXenes_elec_opto_3, MXenes_elec_opto_2}. On the other hand, the biggest challenge in MXenes is keeping their time wise performance at the assumed level. MXenes tend to oxidize due to their low thermodynamic stability. While being nonstoichiometric variants of bulk transition metal carbides and nitrides, they tend to undergo a phase transition in a vacuum from hexagonal into more stable bulk compartments. If the MXene is exposed to an external environment containing oxygen and water, the metallic M element easily reacts with oxide-bearing compounds and M$_x$O$_y$ oxides are being formed. These oxides are more stable than MXene thus, the oxidation is favorable and rapid. All these point to the potential problems with keeping the TE effect of MXene constant in the oxygen-containing environment. However, no specific experimental data are available in this area. Therefore, high-thorough research should be carried out to understand the relationship between MXenes’ oxidation and their TE effect. It is of interest not only from a fundamental point of view but also to practitioners, who put the most effort into bringing the most promising TE materials closer to industrial implementations.

Nevertheless, MXenes have several benefits to make use of them as outstanding TE materials \cite{MXenes_ZHU2023106718}. The high electrical conductivity and the hydrophilicity possessed by MXenes make them suitable for TE applications. This typical feature sets MXenes apart from the conventional TE materials that exhibit n-type conduction like Bi$_2$Te$_3$, Cu$_2$Se, and ZnSe, leading to comparatively lower electrical conductivity. The flexibility of surface termination groups and their intrinsic 2D layered structure of MXenes offer value opportunities for the modification of electronic and thermal properties in a way that is favorable for TE applications. Moreover, the redox activity in the electrochemical processes due to the presence of transition metal atoms on the surface of MXenes proves useful in the construction of a thermoelectrochemical cell. Additionally, in comparison with many TE materials, the better flexibility offered by MXenes grants additional advantages for applications in wearable electronic devices.

During the last decade, MXenes have extended their assistance to the field of thermoelectricity. For TE performance comparison, the standard analyses require determining several quantities such as the Seebeck coefficient, electrical conductivity, thermal conductivity, power factor, and finally, the ZT value. This applies well to simple one-phase structures or assemblies. However, more advanced MXene-based structures are underway to achieve better performance. In particular, nanocomposites based on MXene/carbon, MXene/polymer, and MXene/inorganic composites have been synthesized using different techniques \cite{MXenes_ZHU2023106718}.

Ti$_3$C$_2$T$_x$/PEDOT: PSS nanocomposites have been prepared by Guan \textit{et al} by drop casting a solution of PEDOT: PSS and MXene in water and DMSO \cite{Xin-Guan_TE_2020}. In another study, Jin \textit{et al} prepared composites PEDOT: PSS/MXene/PEG through vertical freezing and vacuum infiltrating method for excellent thermal management and electromagnetic interference shielding performance \cite{JIN2022137599}. Furthermore, a series of composite films consisting of MXenes (Ti$_3$C$_2$T$_x$) and single-walled carbon nanotubes (SWCNT) have been constructed by Ding \textit{et al} \cite{Wenjun-Ding_TE_2020}. These are mostly fine nanocomposite structures, prepared using liquid-phase chemical processes or ultra-precision manipulation at the atomic level such as chemical vapor deposition (CVD). If not deeply understood in terms of material composition and structure, their TE performance cannot be explained as well. Therefore, it is important to employ high-thorough techniques in research on TE materials and devices. The high-resolution scanning and transmission electron microscopy allows looking deeply into the atomic composition and arrangement. Advanced in-situ techniques can further link their TE performance with material parameters.

For instance, the in situ TE measurement systems allow crossing a limit of 20 nm thickness for the film's electrical conductivity, thermal conductivity, or thermoelectricity. Note that the film thickness heavily influences TE performance measurements of lab-scale produced TE films. Going from micro to nanoscale removes considerable uncertainty and allows measuring the TE effect during film deposition inside a thermal evaporator \cite{Insitu_Nguyen_2023}. All these aspects put fine-tuning of MXene-based TE materials into the spotlight.

In this section we present a comprehensive review of the TE properties of MXenes, from the point of view its structural modifications along with functionalization, double transition metal MXenes, creating vdW heterostructures, and using among others MXenes as nanoinclusions or nanocomposites.

\subsection{Surface modification - Functionalization}
\begin{figure}
\centering
\includegraphics[width=0.5\textwidth]{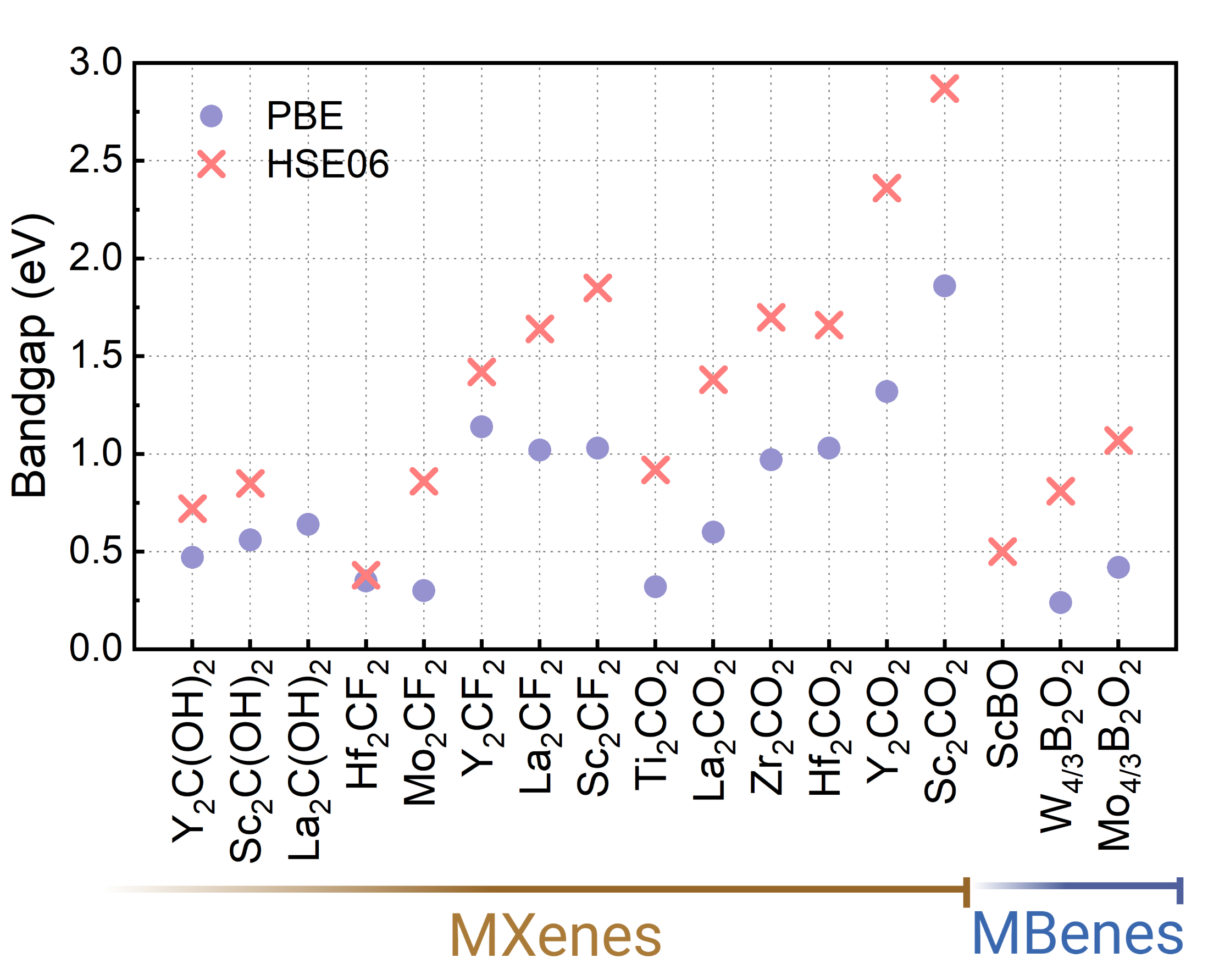}
\captionsetup{justification=raggedright}
\caption{Band  gaps of semiconducting MXenes
and MBenes functionalized with various atoms calculated within the DFT approach assuming  PBE and hybrid
exchange-correlation functionals (HSE06). Reproduced with
permission from Ref. \cite{Abiyyu}}
    \label{MXenes_bandgaps}
\end{figure}

\begin{figure}
\centering
    \includegraphics[width=0.5\textwidth]{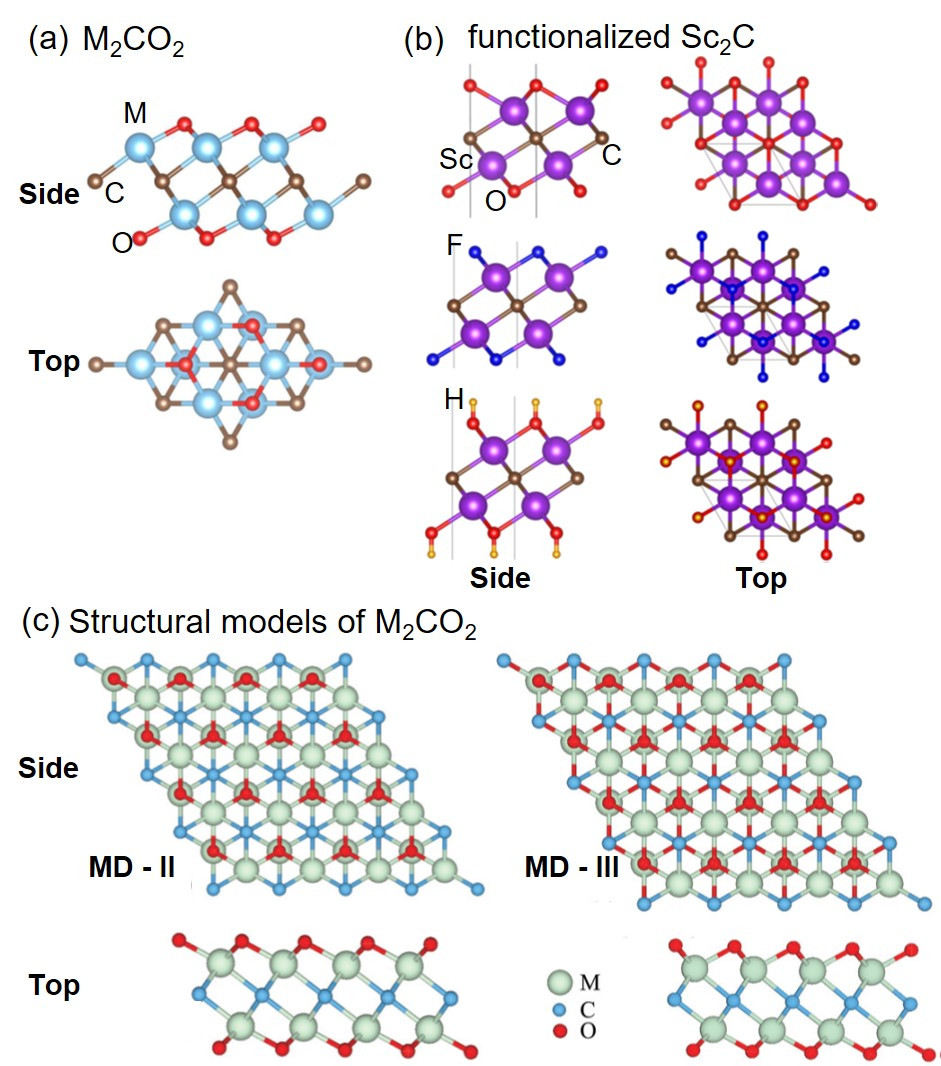}
    \captionsetup{justification=raggedright}
    \caption{ Side and top views of atomic structures of: (a) M$_2$CO$_2$ (M = Ti, Zr, or Hf), Reproduced with permission from Ref. \cite{Gandi_TE_2016}. Copyright 2016 American Chemical Society. (b) Monolayers of Sc$_2$C functionalized by O, F, and OH, Reproduced with permission from Ref. \cite{Kumar_TE_2016}. Copyright 2016 American Physical Society. (c) two different structural models MD-II and MD-III of functionalized M$_2$CO$_2$, Reproduced with permission from Ref. \cite{Sarikurt_TE_2018}. Copyright 2018 Royal Society of Chemistry}
    \label{MXenes_review_articles}
\end{figure}
In general, the MAX phases have been poor TE materials due to their relatively low Seebeck coefficients and high thermal conductivity. However, most of the MXenes behave semiconducting upon suitable surface terminations or functionalization which can lead to promising TE properties. The functionalization of MXenes will lead to band gaps of around 0-3 eV depending on the exchange-correlation functional used \cite{Abiyyu}. Figure \ref{MXenes_bandgaps} presents the semiconducting band gaps of various MXenes with functionalization. There are several approaches to enhance the Seebeck coefficient values of MXenes by band engineering technique. In this sub-section, a detailed review of how to open the band gap of MXenes has been presented. 

 
 









In 2013, Khazeaei \textit{et al} performed first principle calculations and confirmed that Sc$_2$CF$_2$, Sc$_2$C(OH)$_2$, Sc$_2$CO$_2$, Ti$_2$CO$_2$, Zr$_2$CO$_2$, and Hf$_2$CO$_2$ MXenes behave as semiconductors upon functionalization \cite{https:Khazaei_TE_2013}. The electronic structure calculations performed with the aid of Boltzmann transport theory show that Ti$_2$CO$_2$ can attain a very high Seebeck coefficient of about 1140 $\mu VK^{-1}$ at 100 K. Meanwhile, Sc$_2$C(OH)$_2$ gains even larger Seebeck coefficients of around $\approx$ 2000 $\mu VK^{-1}$ at the same temperature with a larger semiconducting gap. These large Seebeck coefficient values can be attributed to the large transition of the density of states (DOS) along with the carrier velocities with functionalization. 


As a follow-up research, Khazaei \textit{et al} also implemented similar calculations to study the TE nature of monolayer and multilayers of M$_2$C (M = Sc, Ti, V, Zr, Nb, Mo, Hf, and Ta) and M$_2$N (M = Ti, Zr, and Hf) MXenes functionalized with F, OH, and O groups \cite{Khazaei_TE_2014}. The calculations predict that the nanosheets of Mo$_2$C MXenes (both monolayer and multilayers) have greater power factors ($S^2\sigma$) than other MXenes at low carrier concentrations with any type of functionalization. Especially, among all 35 different functionalized MXenes, Mo$_2$CF$_2$ has been found to exhibit better power factor values due to its semiconducting gap and its distinctive band shape near the band edges. Meanwhile, V-, Nb-, and Ta-based MXenes exhibit poor TE performance, while Ti-, Zr-, and Hf-based MXenes can be average TE materials.


Nevertheless, it has to be pointed out that the phonon contributions have not been considered to have a total picture of the TE capability of the materials until the work performed by Gandi \textit{et al} \cite{Gandi_TE_2016}. The TE behavior of three MXenes M$_2$CO$_2$ (where M = Ti, Zr, HF) has been analyzed by solving Boltzmann transport equations and taking into account both the electron and phonon contributions. Figure \ref{MXenes_review_articles} illustrates the side and top views of the crystal structure of M$_2$CO$_2$ MXene with the lowest total energy achieved in this work. The phonon calculations have been performed for the first time to estimate the lattice thermal conductivity of MXenes. The lattice thermal conductivity has been found to be lowest in Ti$_2$CO$_2$ and highest in the case of Hf$_2$CO$_2$ respectively, at temperature ranging from 300 K to 700 K. This is because the acoustic phonon branches in the case of Ti$_2$CO$_2$ display strongest dispersions. This results in Ti$_2$CO$_2$ portraying better TE behavior.


A similar kind of DFT study has been also executed to study the TE behavior of Sc$_2$C MXenes functionalized with O, F, and OH \cite{Kumar_TE_2016} using Boltzmann transport theory. The side and top views of the crystal structures of monolayer Sc$_2$C functionalized with O, F, and OH have been shown in Figure \ref{MXenes_review_articles}b. As expected, the metallic nature of the Sc$_2$C has been transformed into semiconducting by opening the band gap with functionalization. The largest Seebeck coefficient has been attained in the case of Sc$_2$CO$_2$ and Sc$_2$CF$_2$ due to larger band gaps than in Sc$_2$C(OH)$_2$. Even though the Seebeck coefficient has been much smaller for OH than O or F functionalization at high temperatures, the power factor remains higher because of high electrical conductivity. This leads Sc$_2$C(OH)$_2$ to be the best TE material as it also had the lowest thermal conductivity due to the largest anharmonic phonon scattering.


It is very clear that the TE properties can be influenced by functionalization. One more possible way to play around with the TE properties is by choosing different structural adsorption sites.  The impact of surface functionalization of MXene monolayers on the TE properties has been analyzed by Sarikurt \textit{et al} considering two different structural models MD-II and MD - III \cite{Sarikurt_TE_2018}. The schematic illustration of side views of the functionalized MXenes M$_2$CO$_2$ and their different structural models are displayed in Figure \ref{MXenes_review_articles}c. The transport properties of oxygen terminated M$_2$CO$_2$ (M = Ti, Zr, Hf, Sc) monolayer MXene crystals have been calculated using DFT and Boltzmann transport equations. The maximum Seebeck coefficient values for the n-type and p-type Ti$_2$CO$_2$, Zr$_2$CO$_2$, Hf$_2$CO$_2$ and Sc$_2$CO$_2$ have been been found to be around 450 - 800 $\mu VK^{-1}$ at room temperature for both the types of structural configurations. It has been demonstrated that the ZT values can be doubled by lowering the crystal symmetry of MXene. Depending on the structural models, the lattice thermal conductivity, Seebeck coefficient, and ZT values differed by around 40 \%. The maximum value of ZT has been computed as 1.1 based on a structural model of MXene. 


It is now well understood that the surface addition of functional groups to the MXenes can influence their band structures. However, the impact of surface functionalization on the thermal conductivity of MXenes has yet to be elucidated in the future. The lattice and electron thermal conductivities of Ti$_2$C MXenes have been explored by the modulation effect of the surface functionalization with the help of Boltzmann transport theory based on DFT \cite{Zhonglu_TE_2018}. The group velocities, Debye temperature, and specific heat of Ti$_2$C MXenes were increased with surface functionalization as observed from the phonon dispersion curves. This results in the increase of the lattice thermal conductivity of Ti$_2$C MXenes by F or OH and enhanced up to 64 \% or 72 \% at 300 K depending on the armchair or zigzag direction. Whereas in the case of O surface functionalization, the lattice thermal conductivity of Ti$_2$C MXenes has been reduced due to the shorter phonon relaxation time of Ti$_2$CO$_2$ MXenes arose from larger Gruneisen parameter and total phase space for three phonon processes. The calculations show that the lattice thermal conductivity of the pristine and functionalized Ti$_2$C MXenes is anisotropic.

A similar approach of surface modification has been applied and synthesized Ti$_3$C$_2$T$_x$ films experimentally. The semiconducting nature of the highly conductive Ti$_3$C$_2$T$_x$ film has been optimized by surface group modification to apply it as TE material \cite{Peng-Liu_TE_2020}. A hydrothermal reaction mechanism has been implemented to substitute -F and -OH terminal groups with -O groups. This surface modification leads to the expansion of the band gap from 0.72 eV to 1.32 eV and enhances the Seebeck coefficient. As a result, the TE power factor has been improved to 44.98 $\mu Wm^{-1}K^{-2}$ at room temperature.

Meantime, the TE nature of Cr-based MXenes Cr$_2$C has been studied using spin-polarized calculations. First principle calculations have been performed on Cr$_2$CT$_2$, Mo$_2$CT$_2$, and W$_2$CT$_2$ \cite{Xiangda-Zou_TE_2018}. Fluoride functionalization of Cr$_2$C and Mo$_2$C modified them into semiconductors with similar band gaps and improved their Seebeck coefficient values to around 1000 $\mu VK^{-1}$ at 100 K irrespective of their low stability. Meanwhile, W$_2$CF$_2$ was still found to be metallic and does not possess any better TE properties.

Very recently, the thermoelectric properties of Yttrium-based MXenes: Y$_2$CT$_2$ (where T = F, Br, OH, H), Y$_2$CClH and Y$_2$CFH have been explored by first principle calculations within the DFT and Boltzmann transport theory \cite{OMUGBE_TE_2022}. The electronic structure calculations show that all the compounds were semiconductors due to functionalization. The ZT values have been found to be high at low temperatures and the highest ZT value 0.97 has been recorded for Y$_2$CH$_2$ with a Seebeck coefficient of 1188 $\mu VK^-{1}$ at 100 K. The calculations demonstrate that Yttrium carbide-based MXenes can be excellent TE materials at low and moderate temperatures.

\subsection{Nanostructuring}
\begin{figure*}
    \centering
 \includegraphics[width=0.7\textwidth]{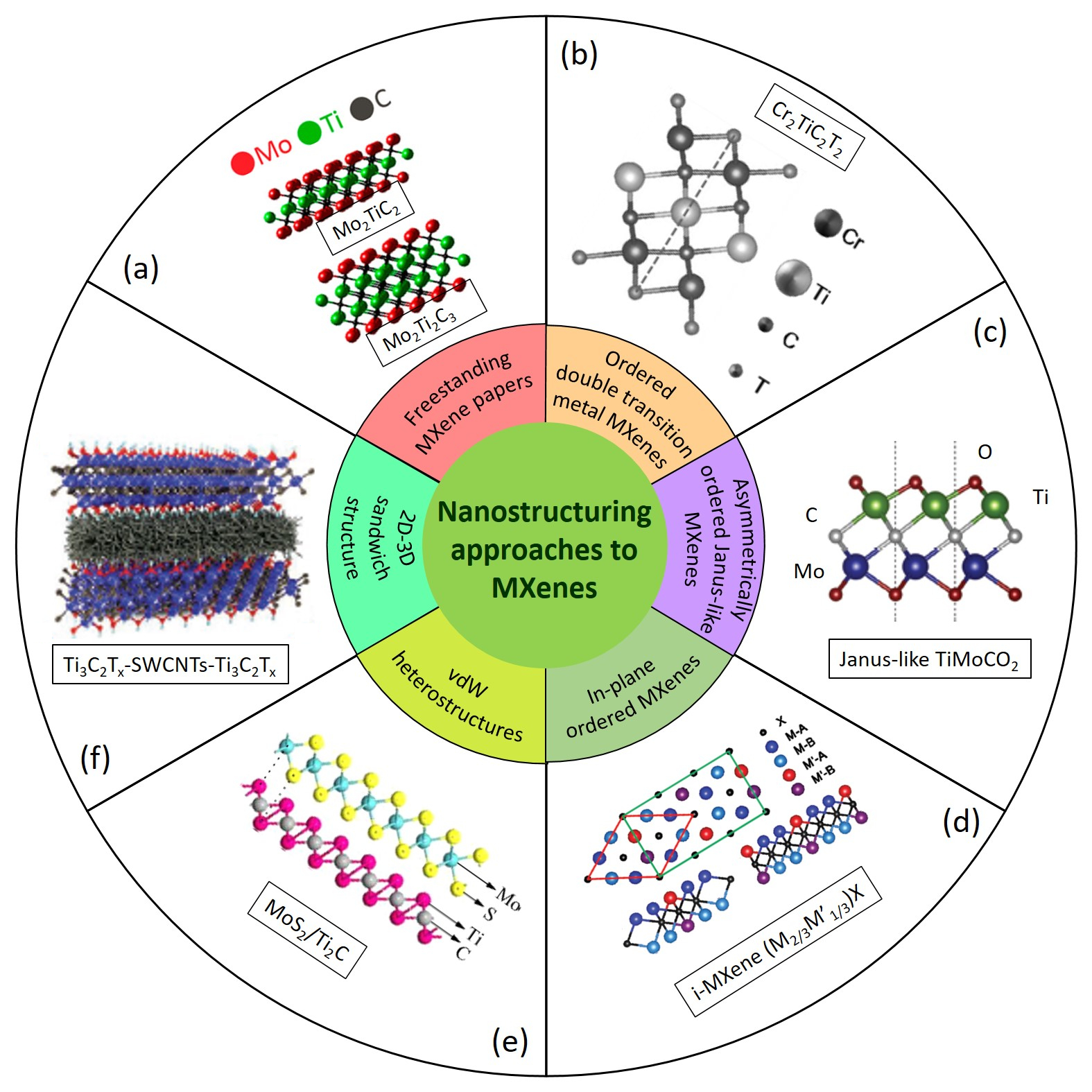}
    \caption{Nanostructuring approaches used for MXenes to enhance the TE properties. (a) Ordered double transition metal MXenes Mo$_2$TiC$_2$ and Mo$_2$Ti$_2$C$_3$, Reproduced with permission from Ref. \cite{Kim_TE_2017}. Copyright 2017 American Chemical Society. (b) Ordered double transition metal MXene Cr$_2$TiC$_2$T$_2$, Reproduced with permission from Ref. \cite{Ziang-Jing_TE_2019}. Copyright 2019 American Chemical Society. (c) Janus-like MXene TiMoCO$_2$, Reproduced with permission from Ref. \cite{Zicong-Wong_TE_2020}. Copyright 2020 Royal Society of Chemistry. (d) In-plane ordered i-MXene (M$_{2/3}$M'$_{1/3}$)$_2$X, Reproduced with permission from Ref. \cite{Qiang-Gao_TE_2020}. Copyright 2020 Royal Society of Chemistry. (e) MoS$_2$/Ti$_2$C heterostructure, Reproduced with permission from Ref. \cite{Chenliang-Li_TE_2018} Copyright 2016 Elsevier. (f) 2D-3D sandwich structure Ti$_3$C$_2$T$_x$–SWCNTs–Ti$_3$C$_2$T$_x$, Reproduced with permission from Ref. \cite{Wenjun-Ding_TE_2020}. Copyright 2020 John Wiley and Sons}
    \label{MXenes_nano_approach}
\end{figure*}

\subsubsection*{Double transition metal MXenes}

The family of MXenes has progressed even further with the breakthrough of double transition metal MXenes \cite{DTM_1}. These kinds of MXenes are classified into ordered and solid-solution MXenes depending on their structure \cite{DTM_2}. The addition of another transition metal to the MXenes and their evolved structure gives more scope for electronic and optical applications.

In the year 2017, Kim \textit{et al} synthesized Mo-based MXenes (Mo$_2$CT$_x$, Mo$_2$TiC$_2$T$_x$, and Mo$_2$Ti$_2$C$_3$T$_x$) in the form of freestanding papers by a simple vacuum filtration method and the TE values have been measured for the first time experimentally \cite{Kim_TE_2017}. The crystallographic orientation of these ordered double transition metal MXenes has been illustrated in Figure \ref{MXenes_nano_approach}a. Mo$_2$TiC$_2$T$_x$ and Mo$_2$Ti$_2$C$_3$T$_x$ have displayed n-type behavior, while the Mo$_2$CT$_x$ exhibited a transition from p-type to n-type character upon annealing. All MXenes reported in this study show a large increase in conductivity values above higher temperatures of 500 K. This is due to the de-intercalation of water and organic molecules and the partial loss of functional groups that reduce the interlayer spacing and improve the contact between the MXene nanosheets. The same is also clearly observed in the annealed MXene papers that are more compact than the pristine MXene papers. Among the three Mo-based MXenes, Mo$_2$TiC$_2$T$_x$ has been found to be better and comparable to conventional TE materials with a superior power factor of 3.09 $\times 10^{-4}$ $Wm^{-1}K^{-2}$ at 803 K. This high value of TE power factor might be due to its peculiar electronic band structure popularly known as "pudding-mold" type that can be seen in high-performance TE materials \cite{pudding_Kuroki, pudding_Mori}. According to the works from Kuroki \textit{et al} \cite{pudding_Kuhas} and Mori \textit{et al} \cite{pudding_Mori}, it has been concluded that the peculiar band shape can be responsible for high Seebeck coefficient values with low resistivity values at the same time. It is well known that the majority of the band shapes will be parabolic, while in contrast to this, the bands appear to be flat and dispersive on the top or bottom near the fermi level in the case of pudding-mold type band structures.

In the meantime, the TE performance of the ordered double MXenes has been also inspected by DFT calculations. The transport properties of ordered double transition metal MXenes Cr$_2$TiC$_2$T$_2$ (T = -OH or -F) (as shown in Figure \ref{MXenes_nano_approach}b) have been analyzed with the aid of spin-polarized DFT calculations using SCAN-rVV10 exchange-correlation functional and  Boltzmann transport theory by Jing \textit{et al} \cite{Ziang-Jing_TE_2019}. The Seebeck coefficients of all these Cr-based ordered double transition metal MXenes with functionalization have been found to be large enough for better TE efficiency. The surface functionalization not only influenced the reduction of phonon thermal conduction of Cr$_2$TiC$_2$(OH)$_2$ to around 6.5 $Wm^{-1}K^{-1}$ but also increased its electron thermal conduction to around 50 $Wm^{-1}K^{-1}$. The ZT values of the hole-doped Cr$_2$TiC$_2$(OH)$_2$ have been reached up to 2.58 at 300 K and 3.0 at 600 K with an exceptional TE efficiency of around 20 \%, implying that these ordered double transition metal MXenes can be excellent candidates for TE applications.

One more work performed on the double transition metal MXenes Ti$_{3-x}$Mo$_x$C$_2$ (x = 0.5, 1, 1.5, 2, 2.5) with -OH/-O/-F functionalization has been investigated theoretically \cite{Shiladitya-Karmakar_TE_2020}. The transport coefficients have been computed especially in the case of semiconducting Ti$_2$MoC$_2$F$_2$, TiMo$_2$C$_2$F$_2$, and    TiMo$_2$C$_2$(OH)$_2$ for both p-type and n-type doping at a carrier concentration of $10^{19} cm^{-3}$. A large value of Seebeck coefficient of around 300 - 400 $\mu VK^{-1}$ was found for the p-type compounds while very low values have been recorded in the case of n-type compounds. Ti$_2$MoC$_2$F$_2$ acquired the highest ZT values than TiMo$_2$C$_2$(OH)$_2$ and TiMo$_2$C$_2$F$_2$. A high ZT value of 3.1 at 800 K and 1.5 at 300 K has been found in the p-type Ti$_2$MoC$_2$F$_2$ resulting from its higher power factor compared to the Ti$_2$Mo$_2$C$_2$. An efficiency of $\sim27 \%$ at 800 K achieved by the p-type Ti$_2$MoC$_2$F$_2$ made it an efficient TE material.

Meanwhile, a multiscale modeling concept has been applied to analyze the asymmetrically ordered Janus-like (2-faced) Ti-Mo based MXene alloys. Ti$_{2(1-x)}$Mo$_{2x}$CO$_2$ (0 < x < 1) alloy systems of 2D Ti$_2$CO$_2$ and Mo$_2$CO$_2$ MXenes have been studied with the aid of multiscale approach of DFT, cluster expansion method, and Monte Carlo simulations \cite{Zicong-Wong_TE_2020}. The top view of the Janus-like ground state TiMoCO$_2$ MXene alloys is shown in Figure \ref{MXenes_nano_approach}c. The asymmetrically ordered Janus-like MXene alloy TiMoCO$_2$ shows high n-type and p-type power factor values of 49.8 $\mu Wcm^{-1}K^{-2}$ and 15 $\mu Wcm^{-1}K^{-2}$ respectively at room temperature and even surpassed other experimentally synthesized Ti-Mo based MXene alloys \cite{Kim_TE_2017}.

A similar kind of theoretical study has been performed on a narrow gap Janus MXene monolayer MoWCO$_2$ \cite{Janus_2}. The TE and mechanical properties have been calculated by taking into account the surface effect and pointing out that these materials can be used for wearable TE devices and sensor applications. The monolayer exhibited power factor values of 6.5 $\times 10^{3}$ $ \mu Wm^{-1}K^{-2}$ for p-type and 1.5 $\times 10^{3}$ $ \mu Wm^{-1}K^{-2}$ for n-type at 700 K. It also revealed a high lattice thermal conductivity of around 308 $Wm^{-1}K^{-1}$ at room temperature. As a result, very poor ZT values of around 0.04 for p-type and 0.01 for n-type at 700 K have been recorded for this monolayer. The large thermal conductivity values can be attributed to the long phonon lifetime. To enhance the ZT values, the boundary scattering mechanism has been considered to reduce the lattice thermal conductivity. The TE parameters have been estimated as a function of the width of the nanoribbon of the 2D Janus monolayer to figure out the effect of the material on the TE performance. The lattice thermal conductivity has been reduced to 65 $Wm^{-1}K^{-1}$ at 300 K for 1 $\mu$m width of the nanoribbon. Accordingly, the ZT has been improved to 0.33 for p-type and 0.08 for n-type at 700 K by assuming the width of the nanoribbon to be 10 nm.

2D in-plane ordered MXenes (i-MXenes) have been studied using DFT calculations to investigate their magnetic nature, TE properties, and several other multifunctional applications \cite{Qiang-Gao_TE_2020}. The top views and side views in both hexagonal and rectangular lattice of i-MXenes (M$_{2/3}$M'$_{1/3}$)$_2$X have been illustrated in the Figure \ref{MXenes_nano_approach}d. The electronic properties of the i-MXenes in the rectangular and hexagonal geometries revealed excellent transport properties. Even though most of the i-MXenes turned out to be metallic, a few semiconducting compounds have also been found. In particular, (Sc$_{2/3}$Cd$_{1/3}$)C and (Sc$_{2/3}$Hg$_{1/3}$)C portray larger band gaps, implying they exhibit higher Seebeck coefficient values. The semiclassical transport calculations performed on these two i-MXenes show that (Sc$_{2/3}$Cd$_{1/3}$)C and (Sc$_{2/3}$Hg$_{1/3}$)C can be good TE materials with large Seebeck coefficients of around 1000 $\mu VK^{-1}$ at room temperature. The calculations propose that i-MXenes can be utilized for the most promising future spintronic applications.

\subsubsection*{Stacking different types of layers - vdW heterostructures}
Another way to customize the properties of 2D materials is to design optimal heterostructures to enhance TE conversion efficiency. Stacking different 2D materials into van der Waals (vdW) heterostructures is an efficient route to manipulate the electronic structure and optimize the TE performance of the materials.

The first principle calculations performed on the semiconducting MXenes (Ti$_2$CO$_2$, Zr$_2$CO$_2$, and Hf$_2$CO$_2$) show that they possess high thermal conductivities at room temperature irrespective of their TE capability\cite{https:Khazaei_TE_2013}. While the monolayer MoS$_2$ from the family of TMDs is known to be a potential material for its electronic and optoelectronic applications. Nonetheless, it has the disadvantage of larger effective mass and low carrier mobility that restricts it from being promising. Gandi \textit{et al} implemented the idea of vdW heterostructures formed by MXenes with TMDs and calculated the thermal conductivity values \cite{Gandi_TE_2016/2017}. The TE response in the heterostructures of Mo$_2$CO$_2$ (where M = Ti, Zr, HF) with TMDs monolayers has been investigated. The calculations show that the band gap increases with the mass of the transition metal atom. Due to the weak vdW bonding between the 2D layers, low frequency optical phonons advance to the thermal transport and the scattering of the acoustic phonon scattering has been enhanced. As a result, it has been found that the reduction of the heterostructure's lattice thermal conductivities has not been significant compared to the bare MXenes. It has been suggested that the superlattice design approach of thermoelectrics can be even more effective for 2D vdW materials with intercalation to reduce thermal conductivity. The Ti$_2$CO$_2$-MoS$_2$ has been found to be an efficient TE material with a larger ZT value. 

It has been clear now that the theoretical study performed on the functionalized Ti$_2$C/MoS$_2$ did not yield fruitful results. Despite the contrary, an almost similar type of vdW heterostructures has been constructed with bare MXene Ti$_2$C without surface functionalization and MoS$_2$ \cite{Chenliang-Li_TE_2018}. The heterostructure of MoS$_2$/Ti$_2$C is shown in Figure \ref{MXenes_nano_approach}e. The TE properties of MoS$_2$/Ti$_2$C heterostructures have been investigated using the first principle calculations by considering the impact of external strain and electric field. The application of the external field to this heterostructure creates a band gap. The bands become more dispersive, increasing the conductivity with the application of tensile strain. Meanwhile, the energy bands become more flat leading to the reduction in conductivity with the combination of tensile strain and vertical electric field. The change in the electronic properties impacted the TE performance of MoS$_2$/Ti$_2$C heterostructure. Nevertheless, the TE properties can be improved by coupling tensile strain and electric field. 

At the same time, the strategy of vertical stacking of the 2D materials SiS and MXenes into layered vdW heterostructures has also been implemented and studied by first principle calculations based on DFT for different applications \cite{Aqsa-Abid_TE_2021}. SiS monolayers possess an indirect band gap with good mechanical and thermal stability. The individual capabilities of SiS and the MXenes paved the way to construct heterostructures. It has been mentioned that the pristine monolayers SiS, Ti$_2$CO$_2$, Zr$_2$CO$_2$, and Hf$_2$CO$_2$ were intrinsic semiconductors and exhibit the Seebeck coefficient values of -2356, 1751, 640 and 1640 $\mu VK^{-1}$ at 300 K and -980, 632, 235 and 584 $\mu VK^{-1}$ at 800 K respectively. However, when it comes to the vdW heterostructures, the band structure calculations reveal that SiS-Ti$_2$CO$_2$ turns out metallic where as SiS-Zr$_2$CO$_2$ and SiS-Hf$_2$CO$_2$ remained low band gap semiconductors with indirect band gaps. The least effective masses of SiS-Ti$_2$CO$_2$ and SiS-Zr$_2$CO$_2$ made them possess high carrier mobility. The transport properties of these vdW heterostructures show a decreasing trend with the temperature elevation. Among the three vdW heterostructures, SiS-Hf$_2$CO$_2$ exhibits greater power factor values because of its higher sensitivity towards carrier concentration.

\subsubsection*{Stacking 2D layers on different dimensions}

A heterostructure can be designed not only by 2D materials but also by stacking layers of different dimensions. Carbon nanotubes (CNTs) with 3D porous structures are very well established with amendable network morphology which can be utilized to optimize the carrier mobility and thermal conductivity. Combining different components can be highly advantageous as the mismatch can lead to a charge carrier filtering effect and significantly advance the properties. A 2D-3D sandwich structure Ti$_3$C$_2$T$_x$–SWCNTs–Ti$_3$C$_2$T$_x$ has been constructed with 2D MXene and 1D SWCNT (single-walled carbon nanotube) films to enhance the TE performance \cite{Wenjun-Ding_TE_2020}. The schematic diagram of one of the different combinations of the heterostructure designs is illustrated in Figure \ref{MXenes_nano_approach}f. The higher carrier concentration achieved by this sandwich structure increases the electrical conductivity to 750.9 $Scm^{-1}$. The effective double energy barriers that cause the low-energy carriers scattering enhanced the Seebeck coefficient to -32.2 $\mu VK^{-1}$. This results in the increase of power factor to 77.9 $\mu Wm^{-1}K^{-2}$ which is 25 times higher than that of Ti$_3$C$_2$T$_x$ film. 

Very recently, Jiacheng Wei \textit{et al} synthesized 3D hollow structured 1D-SWCNT/2D-MXene films as p-type TE materials for the first time \cite{Jiacheng-Wei_TE_2022}. The energy filtering effect at the SWCNT/MXene interface improves the Seebeck coefficient of the composite by two times in comparison to SWCNT. This is because the holes in SWCNT occupied by the electrons from MXenes will lead to the reduction of hole concentration and increase the Seebeck coefficient. The 3D hollow structures formed in the composite films assist in decreasing thermal conductivity. The contact junctions that originated between the SWCNT and MXene in the composite also promote electron transfer. This results in the enhancement of ZT by 20 times higher than that of SWCNTs. A TE device has been constructed with 10-SWCNT/MXene-10 legs and it produced a maximum output power of 1.54 $\mu W$ at a temperature difference of 117.3 K.

\subsubsection*{Nanocomposites/Nanoinclusions}

Several other aspects can influence transport properties. In this subsection, we will discuss different ideas that have been implemented to expand the horizon of MXenes to excel as TE materials.

The Ti$_3$C$_2$T$_x$ MXene showed its impact on enhancing the TE efficiency of the most familiar TE  material Bi$_2$Te$_{2.7}$Se$_{0.3}$ compound by incorporating them as nanocomposites \cite{Dewei-ZHANG_TE_2022}. At 380 K, for Bi$_2$Te$_{2.7}$Se$_{0.3}$/0.1 wt \% Ti$_3$C$_2$T$_x$, the electrical resistivity has been reduced by 43 \% which leads to a power factor of 1.49 $Wmm^{-1}K^{-2}$. The addition of nanocomposites also increases the phonon scattering resulting in the decrease of lattice thermal conductivity from 0.77 to 0.41 $Wm^{-1}K^{-1}$ and enhances the ZT value to 0.68 at 380 K. The enhancement of the efficiency has been found to be 48 \% when compared with the pristine sample.

Similarly, the Ti$_3$C$_2$T$_x$ MXene has also been used to improve the TE power generation by incorporating it in the (Bi, Sb)$_2$Te$_3$ for a wide temperature range \cite{BST-Ti3C2Tx}. Highly conductive MXene Ti$_3$C$_2$T$_x$ as a second phase has been homogeneously dispersed in p-type Bi$_{0.4}$Sb$_{1.6}$Te$_3$ (BST) matrix to fabricate Ti$_3$C$_2$T$_x$/BST composite. An increase in power factor along with the reduction of lattice thermal conductivity has been found in the composite with Ti$_3$C$_2$T$_x$ functionalized with oxygen. As a result, an average ZT of $\approx$ 1.23 has been obtained in the temperature range of 300 to 475 K. The composite has displayed a TE conversion efficiency of 7.8 \% under a temperature difference of 273 K. 

Guan \textit{et al} incorporated an n-type MXene Ti$_3$C$_2$T$_x$ into p-type poly (3,4-ethylene dioxythiophene): polystyrene sulfonate (PEDOT: PSS) and investigated the TE performance \cite{Xin-Guan_TE_2020}. With the integration of MXene (wt \%), there has been an improvement of the Seebeck coefficient from 23 to 57.3 $\mu VK^{-1}$ and power factor from 44.1 to 155 $\mu Wm^{-1}K^{-2}$ to the composite. This enhancement has been dedicated to the energy filtering of charge carriers by the internal electric field that emerged at the interface between Ti$_3$C$_2$T$_x$ MXene and PEDOT: PSS. 
\begin{figure}
    \centering
    \includegraphics[width=0.4\textwidth]{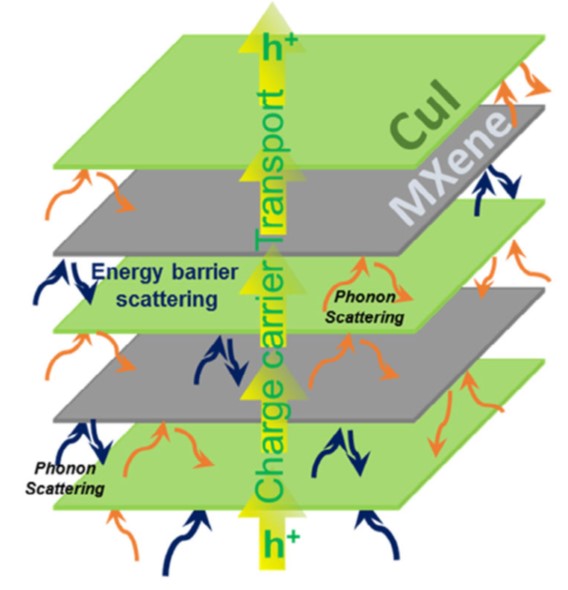}
    \caption{The effect of interlaced architecture for energy barrier scattering between CuI and Ti$_3$C$_2$. Reproduced with permission from Ref. \cite{Vaithinathan-Karthikeyan_TE_2022}. Copyright 2022 John Wiley and Sons}
    \label{CuI-MXene}
\end{figure}

It is well known that nanocomposites or nanoinclusions can reduce the thermal conductivity values, improving the TE behavior of the materials. Ti$_3$C$_2$ MXene nanoinclusions have enhanced the TE properties in the hierarchically interlaced novel 2D copper iodide nanoflakes \cite{Vaithinathan-Karthikeyan_TE_2022}. The interlaced architecture of CuI/Ti$_3$C$_2$ and the mechanism of the charge carrier transport energy barrier scattering and the phonon scattering are portrayed in Figure \ref{CuI-MXene}. The electrical conductivity of the interlaced structure of the CuI/Ti$_3$C$_2$ composite has been raised over almost two orders due to the significant charge transport mechanisms. Meanwhile, the composite's thermal conductivity has been extremely reduced as the interlaced composite allows the interfacial energy barrier scattering of mid- and high- frequency phonons. With a 5 vol \% of Ti$_3$C$_2$ nanoinclusion in the CuI matrix, an impressive power factor of 225 $\mu Wm^{-1}K^{-2}$ and a ZT value of 0.48 has been obtained for CuI/Ti$_3$C$_2$ composite around 550 K.

\subsection{Potential applications of MXenes-based materials}

The TE efficiency possessed by MXenes led them to take advantage of building devices like terahertz detectors, and TE generators and also use them in photothermoelectric applications.

The strong photon absorption, fast carrier relaxation, and weak electron-phonon coupling of a TE material can show its ability to be a potential terahertz (THz) detector. DFT calculations performed on stacked Ti$_3$C$_2$ flakes show that their ZT values were similar to that of carbon nanotube films \cite{Y.I.Jhon_TE_2018}. The TE nature of stacked Ti$_3$C$_2$ flakes with -OH and -O surface functionalizations has been examined by designing a device system with a channel and electrode regions and then utilizing nonequilibrium Green's function formalism to study the electronic and phonon transport properties. Due to the decent TE performance and magnificent THz absorption, it has been suggested that these MXenes can excel in THz detection applications. It has been concluded that n-type doping increased the ZT value of Ti$_3$C$_2$ to 0.112 by a Fermi energy shift.
 
A thin-film TE generator (TFTEG) based on Mo-based TE materials has been designed and analyzed \cite{Taeho-Park_TE_2021}. The TE performance of spin-coated p-type Mo$_2$C and n-type Mo$_2$Ti$_2$C$_3$ thin films at room temperature has been investigated. The in-plane thermal conductivity values have been found to be 0.37 and 0.45 $Wm^{-1}K^{-1}$ for the p-Mo$_2$C and n-Mo$_2$Ti$_2$C$_3$ thin films, respectively. The Seebeck coefficient values have been recorded as 308 and -25 $\mu VK^-{1}$ resulting in ZT values of 1.7 $\times 10^{-5}$ and 2.6 $\times 10^{-4}$ for the p-Mo$_2$C and n-Mo$_2$Ti$_2$C$_3$ thin films correspondingly. It has to be noted that the Seebeck coefficient values of these spin-coated films have been about 20 times higher than the Mo$_2$C papers fabricated by \textit{Kim et al} and also more suitable for constructing TFTEG \cite{Kim_TE_2017}. A Seebeck voltage of 399.9 mV with an output power of 6 mW$cm^{-2}$ at $\Delta$T of 5.4 K has been generated for the MXene TFTEG designed with 200-pn modules. 

\begin{figure}
    \centering
    \includegraphics[width=0.5\textwidth]{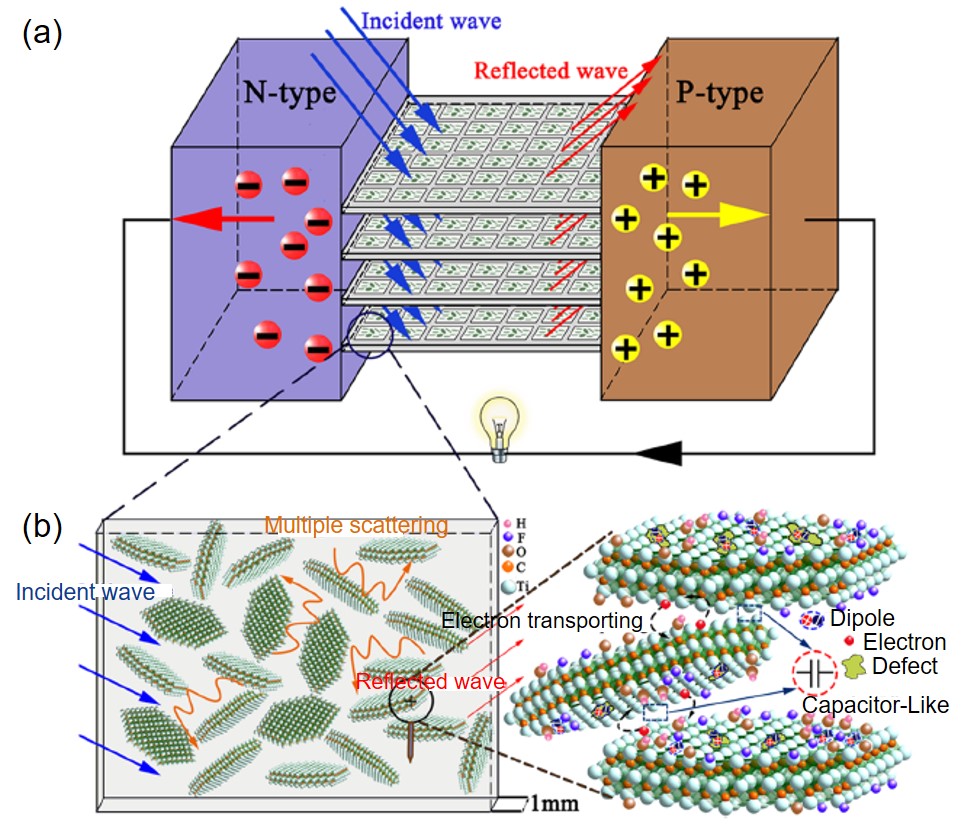}
    \caption{(a) Thermoelectric generator driven by electromagnetic wave. (b) Relevant operational mechanism. Reproduced with permission from Ref. \cite{Peng-He_TE_2019}. Copyright 2019 American Chemical Society}
    \label{MXenes-TE-generator}
\end{figure}

 Ti$_3$C$_2$T$_x$ nanosheets were synthesized by the modified Gogotsi's method \cite{Peng-He_TE_2019}. An electromagnetic driven thermoelectric generator was designed based on Ti$_3$C$_2$T$_x$ nanosheets and displayed in Figure \ref{MXenes-TE-generator}. This prototype successfully converted electromagnetic energy into thermal energy. 

MXenes can also serve the purpose of photothermoelectric applications. The photothermoelectric effect (PTE) is quite similar to the Seebeck effect where a temperature gradient can be achieved by absorbing light on a TE material. In this work, it has been demonstrated how to convert the changes in temperature to electric signals through preferential diffusion of cations under thermal gradient with the help of MXene-based subnanometer ion channels \cite{Seunghyun-Hong_TE_2020}. Lamellar Ti$_3$C$_2$T$_x$ MXene membranes have been investigated as a bionic thermosensitive platform by employing subnanometer confine ion conductors. A photothermoelectric ionic response evaluated from the ionic Seebeck coefficient has been found to be about 1 $mVK^{-1}$ under one sun illumination and was comparable to biological thermosensory channels. 

Meanwhile, the TE ability of Mo$_2$TiC$_2$T$_x$ and Nb$_2$CT$_x$ MXenes have been composed into n-type and p-type junctions, respectively, by organic molecule intercalation and thermal treatment to construct a TE nanogenerator \cite{MXene_TE_nanogenerator}. After optimizing, Mo$_2$TiC$_2$T$_x$ and Nb$_2$CT$_x$ MXenes exhibit TE power factor values of 13.26 $\mu Wm^{-1}K^{-2}$ and 11.06 $\mu Wm^{-1}K^{-2}$ at room temperature, respectively. And using Ti$_3$C$_2$T$_x$ as an electrical contact and Mo$_2$TiC$_2$T$_x$ and Nb$_2$CT$_x$ as n-type legs and p-type legs, respectively, an all-MXene TE nanogenerator has been fabricated. An output power of 33.5 nW has been obtained for a nanogenerator utilizing 20 p-n pairs with a thermal gradient of 30 K. 

\begin{figure*}
    \centering  \includegraphics[width=0.75\textwidth]{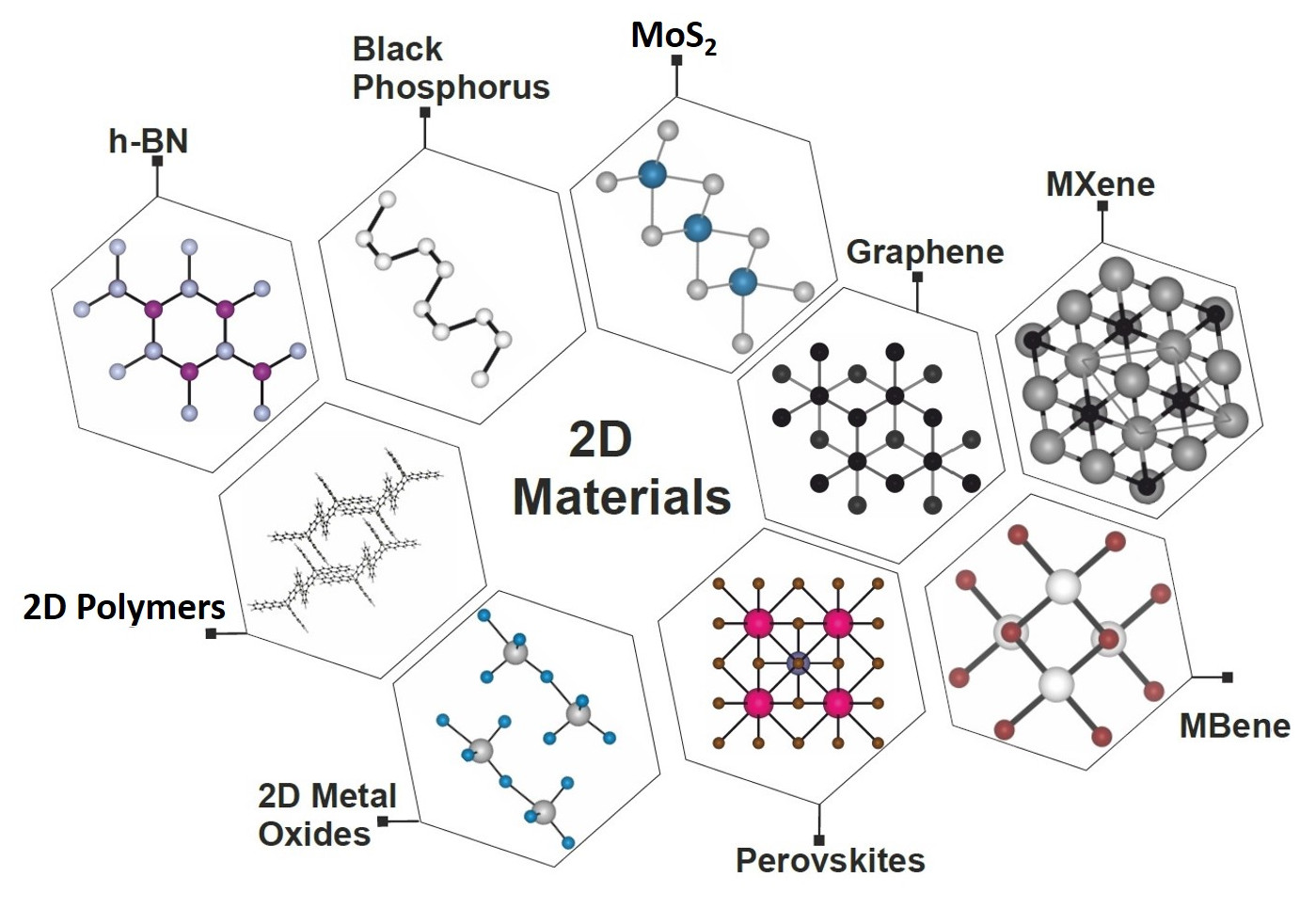}
    \caption{Different types of 2D materials discussed in this paper.}
    \label{2Dmaterials}
\end{figure*} 
An alternative way to overcome the limitations of traditional TE devices and to convert low grade waste into electrical energy is the fabrication of thermocells which are also known as thermoelectrochemical cells or thermogalvanic cells. These thermocells can convert body heat energy into electrical energy. A square shaped thermocell composed of a ternary composite electrode consisting of  Ti$_3$C$_2$T$_x$, SWCNT, and PANI (T-S-P) resulted in better thermoelectrochemical performance in comparison with the widely adopted noble platinum electrodes and displayed a great potential in harvesting or converting body heat into electrical energy \cite{T-S-P}.

A bimodal sensor has been constructed using Ti$_3$C$_2$T$_x$-graphene nanosheet through printing technique that can be used for multiple sensing capabilities, health monitoring, and wearable applications \cite{Bimodal_MXene_Graphene}. A heterogeneous asynchronous photo-thermal-electric (PTE) conversion system based on PEDOT: PSS/MXene has been constructed that uses sunlight as a source \cite{JIN2022137599}. MXene/CCS@CF has been constructed using a high fire safety cotton fabric coated with MXene nanosheet and carboxymethyl chitosan (CCS) \cite{MXenes-CCS}. This MXene/CCS@CF can be utilized for temperature sensing, fire-warning, piezoresistivity, and Joule heating performance. In a later work, with the combination of polydopamine-decorated MXene (P-MXene) and biomass-derived phase change materials (PCM), a solar-thermal electricity conversion device was developed \cite{P-MXene}.

\section{TE properties of other 2D materials}

In this section, selected types of 2D materials (see  Figure \ref{2Dmaterials}) exhibiting promising TE properties are discussed. Notably, their recent progress in thermometric properties has been well reviewed by Delong Li \textit{et al} \cite{2DTEreview_Li_2020}. In this work, we emphasize selected aspects such as structural anisotropy exhibited by the layered materials and layered dependent properties that could affect the TE response. These issues are crucial concerning other newly discovered families of 2D materials - MBenes presented in a further section.
\subsection*{VdW materials with layered dependent TE properties }
\begin{figure}[ht]
    \centering
    \includegraphics[width=0.45\textwidth]{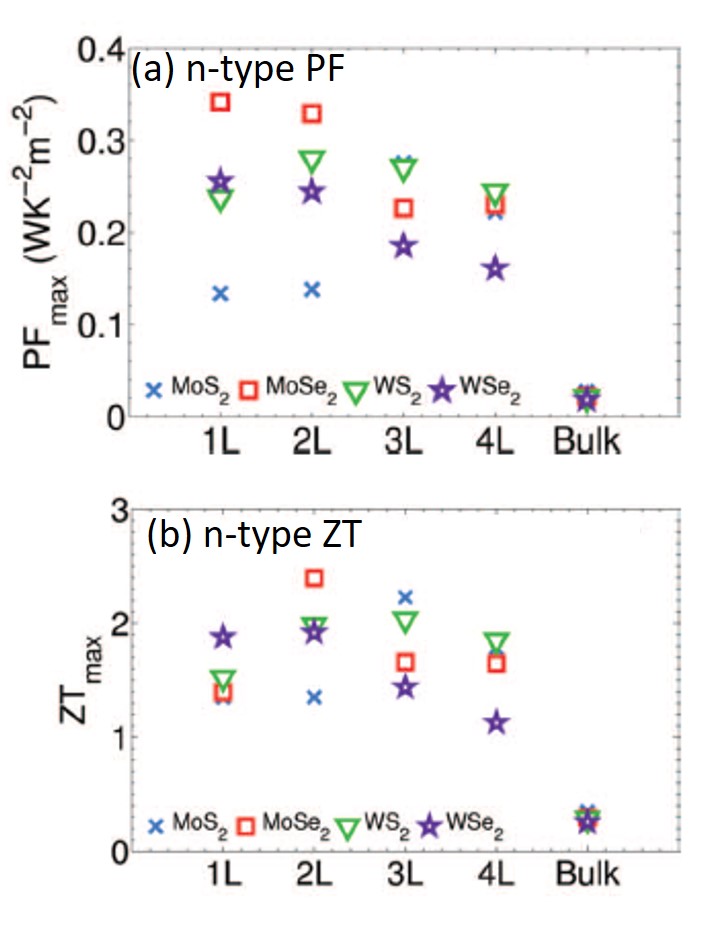}
    \caption{Layered dependent TE properties of TMDs. Maximum (a) n-type power factor values and (b) n-type ZT values attained for 1 layer (blue), 2 layers (red), 3 layers (green), 4 layers (purple), and bulk (black) MoS$_2$, MoSe$_2$, WS$_2$, WSe$_2$ at 300 K, Reproduced with permissions from Ref. \cite{Layered}. Copyright 2014 AIP publishing}
    \label{ZT-TMDs}
\end{figure}

One of the most widely studied layered materials are transition metal dichalcogenides (TMDs), with the general formula of MX$_2$, where M stands for transition metal (Mo, W, etc.), and X is chalcogen atom (X=S, Se, Te). In particular, TMDs are semiconducting materials with a band gap depending on their thickness ranging from 1-2 eV \cite{TMDC_1}. For instance, the monolayer of MoS$_2$ exhibits an indirect semiconducting band gap of 1.9 eV, while its bulk counterpart shows a direct band gap of 1.2 eV. These materials demonstrate excellent performance as TE materials  \cite{2DTEreview_Li_2020, 2D_1_Zhou, 2D_2_Zhang}, including high Seebeck coefficient, electrical conductivity, and reasonably low thermal conductivity \cite{TMDC_TE_1, TMDC_TE_2, TMDC_TE_3}.  The recent report has predicted that the ZT values are very promising in comparison to the bulk materials as the electrical conductivity values depend on the layer thickness \cite{TMDC_TE_3}. The maximum
value of ZT correlates with the sharpest turn-on of the density of modes (DOM)  at the band edges. As the number of layers increases, more bands appear due to the k$_z$ dispersion, and the valleys in the vicinity of the Fermi level started to be nearly degenerated, resulting in enhancement of the DOM, and hence, the enlargement of ZT values (see Figure \ref{ZT-TMDs}). By tuning the valley degeneracy via the thickness of the structure, the separate control of the Seebeck coefficient and electrical conductivity can be reached, pointing to the role of the interlayer interaction in TE properties \cite{TMDC_TE_3}. On the other hand, the Seebeck coefficient is nearly independent of the layer thickness in TMDs \cite{}. As opposed to TMDs, the BP shows strong layer dependence on the Seebeck coefficient. In particular, a monolayer of BP possesses a Seebeck coefficient value of 335 $\mu VK^{-1}$, while for the layered BP (bulk) a significant increase to 510 $\mu VK^{-1}$ at room temperature is reported \cite{BP_TE_Seebeck_1, BP_TE_Seebeck_2}. Hence, a wide range of applications including nanomechanical resonators, TE devices, and motion sensors can be relevant to this material.

On the other hand, the proximity of the different types of layers might also affect the TE properties. In particular, graphene exhibits a high electrical conductivity of about 10\textsuperscript{6} Scm\textsuperscript{-1} at room temperature due to high electron mobility \cite{Graphene_TE-electricalconductivity}. However, the main drawbacks of graphene are a low Seebeck coefficient of 80 mVK\textsuperscript{-1} and a very high thermal conductivity of around 5000 WmK\textsuperscript{-1} at room temperature \cite{Graphene_TE_Seebeck1, Graphene_TE_Seebeck2, Graphene_TE_Seebeck3, Graphene_TE-thermalconductivity}, which prevents graphene to be used as an efficient TE material. Notably, integration of the graphene with other 2D materials such as h-BN enormously enhanced the power factor of graphene \cite{h-BN_TE_1}. In addition, using h-BN as a substrate of graphene or MoS$_2$, the TE performance has been significantly enhanced \cite{h-BN/Graphene,h-BN/MoS2}. For the latter, the TE power factor has been improved by two orders of magnitude in the MoS$_2$/h-BN device \cite{h-BN/MoS2}. The encapsulation of the BP by the h-BN induced changes in the vibrational properties of the latter \cite{Birowska_2019}. Hence, any modification, such as using a different substrate, or creating a vdW heterostructure might affect the TE properties. Moreover, researchers can make TMDs suitable for potential TE generators or coolers by modifying their TE nature. Even though TMDs keep their low thermal conductivity values, the theoretical calculations show that they are still much higher than other conventional TE materials \cite{TMDC_TE_kappa_1}\cite{TMDC_TE_kappa_2}\cite{TMDC_TE_kappa_3}.

\subsection*{Structural anisotropy}\label{sec:anisotropy}
Anisotropy is an additional factor that plays an important role in enhancing the ZT values in TE materials.  Due to the different types of bonding, namely covalent within the layer and weak vdW between adjacent layers, the layered materials naturally host structural anisotropy. The directional dependent TE properties are commonly examined considering in-plane ($\parallel$) and out-of-plane perpendicular  ($\perp$) directions with respect to the basal plane of the layer. Hence, in many types of layered materials direction-dependent TE properties have been investigated, like in hexagonal  BaTiS$_3$  \cite{Paudel2020-ox}, or in In$_2$Te$_5$ \cite{In2Te5}. In the latter, the thermopower measurements demonstrated that the lattice thermal conductivity and electrical resistivity measured along $\perp$ and $\parallel$ significantly differ. Most of the widely studied layered materials such as graphene, MXenes, and TMDs exhibit hexagonal symmetry within the plane for which the directional in-plane properties are isotropic. However, for the systems exhibiting an orthorhombic symmetry, the in-plane anisotropy exists like in monolayer of BP, SnSe as well as newly discovered 2D MBenes.

\begin{figure*}
    \centering
    \includegraphics[width=1.0\textwidth]{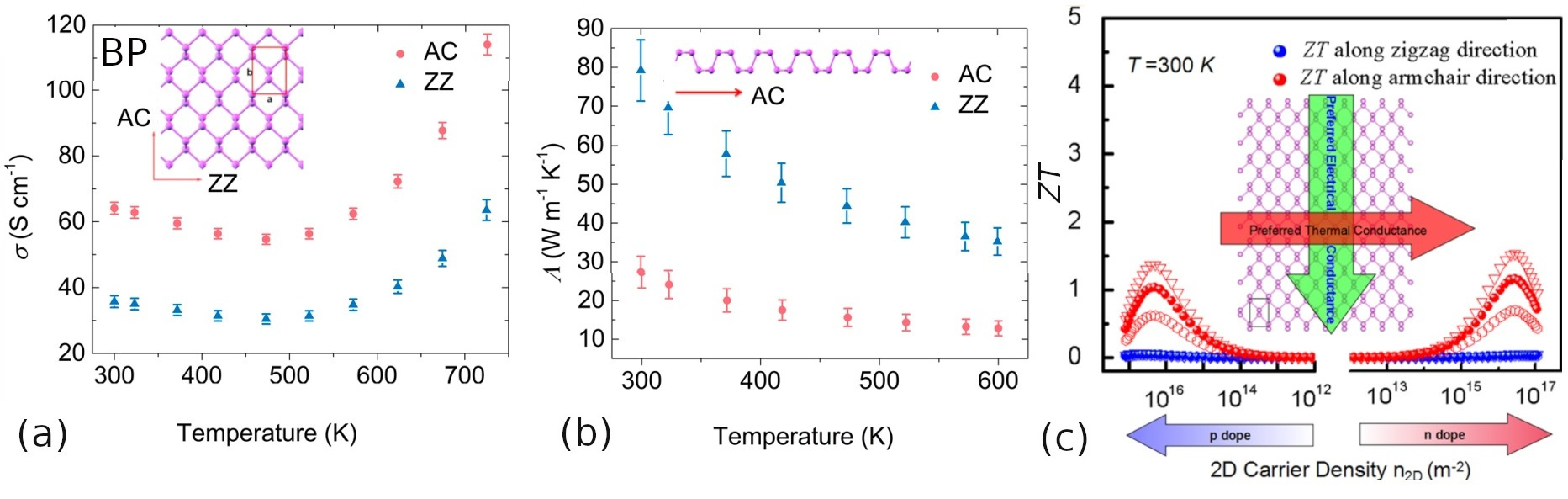}
    \caption{The measured (a) electrical and (b) thermal conductivities as a function of temperature for BP crystals along the armchair (AC) and zigzag (ZZ) directions, Reproduced with permissions from Ref. \cite{BP_anisotropy_zigzag_armchair}. Copyright 2019 IOP Publishing Ltd. The insets in (a,b) show the atomic structure of the monolayer of BP. (c) Predicted the ZT values with respect to the doping density [$m^{-2}$] under T=300K for a monolayer of BP, Reproduced with permission from Ref. \cite{Fei2014}. Copyright 2014 American Chemical Society}
    \label{BP-anisotropy}
\end{figure*}

The BP is an extremely important example, which possesses the remarkable in-plane structural anisotropy originated from the orthorhombic crystal structure (space group $Pnma$) \cite{BP_1, BP_2}. As it is presented in the inset of Figure \ref{BP-anisotropy}(a), each Phosphorus atom is covalently bonded with its three neighbors forming puckered honeycomb layer, with two orthogonal directions along a zigzag (ZZ) and armchair (AC) axes. This strong in-plane anisotropy is reflected in its various properties \cite{BP_TE_anisotropy_1, BP_TE_anisotropy_2,Birowska_2019,BP_TE_anisotropy_3, BP_TE_anisotropy_4, BP_TE_anisotropy_5, BP_TE_anisotropy_6}.  In particular,  the electric and thermal conductance possess strong in-plane anisotropies, resulting in large ZT along the armchair direction in comparison to the zigzag axis \cite{Fei2014}. These results were confirmed by the ZEM-3 and time-domain thermoreflectance (TDMR) measurements of electrical and thermal conductivities, respectively as shown in Figure \ref{BP-anisotropy}(a,b) \cite{BP_anisotropy_zigzag_armchair}. Namely, the ZT along the armchair direction is around 5.5 times larger than along the zigzag direction. Notably, electrical and thermal conductances are not only spatially anisotropic but what is even more pronounced, they possess respectively orthogonal preferred conducting directions (see inset of Figure \ref{BP-anisotropy}(c)), resulting in large $\sigma/\kappa$ ratio \cite{Fei2014}, whereas the Seebeck coefficient exhibits almost isotropic behavior.

  The other layered materials that exhibit the orthorhombic crystal structure such  SnSe and SnS,  also show significant in-plane anisotropic electrical conductivity and Seebeck coefficients. Consequently, the ZT reaches the highest optimal value along the $b$ axis equal to 2.6 at 923 K, and the lowest along the $a$ axis equal to 0.8 \cite{Zhao2014,PhysRevB.92.115202}. Moreover, the recent report has demonstrated that the anisotropy of electronic transport properties of the n-type SnSe can be greatly enhanced by inducing a pressure \cite{Cao2022}. It is worth mentioning, that the SnSe material is considered nowadays as an outstanding TE material, mostly due to the intrinsically low thermal conductivity reaching the value 0.23 $Wm^{-1}K^{-1}$ at 973K, originating from the ionic-potential anharmonicity \cite{Li2015}.   Meanwhile, another 2D material, silicene, with a high specific surface area possesses very good optical properties and excellent electronic properties which can be further extended to remarkable TE applications \cite{Silicene_1, Silicene_2}. The recent report has demonstrated that its hydrogenation can greatly enhance the ZT value of silicene reaching the value of 2.2  along the $y$ direction at 800 K and under the $n$-type doping \cite{Silicene_TE}. In addition, in the newly derived form of silicene - penta-silicene, a  directional-dependent thermal lattice conductivity has also been observed \cite{Gao2020}.

In light of the possible applications, it is well established that anisotropic electrical properties affect device performance, while the anisotropic thermal transport properties are related to the thermomechanical reliability of functional devices \cite{BP_TE_anisotropy_1}. In particular, alternative flexible substrates and active layer materials with enhanced thermal conductivity are needed to prevent thermomechanical damage, which is an important challenge in building flexible devices based on 2D materials.

\subsection*{TE properties of other 2D materials}

The other 2D materials, such as  2D perovskites are also being considered as potential TE applications due to their excellent electronic, mechanical, and TE properties \cite{perovskites_1, perovskites_2}. Moreover, in comparison to the 3D perovskites, the 2D perovskites are known to be eco-friendly with low toxicity \cite{perovskites_3, perovskites_4, perovskites_5}. Also, in the case of 2D perovskites, the enhanced electron-phonon coupling along with the acoustic impedance mismatch between the organic and inorganic layers lead to a highly anisotropic thermal dissipation that is advantageous to the TE devices \cite{perovskites_6, perovskites_8}. The 2D perovskites possess the TE applications mostly in their oxides and hybrid versions \cite{perovksites_7}. Compared to conventional TE materials, hybrid perovskite materials can be produced by low energy cost procedures and used for flexible TE devices \cite{perovksites_7}. A giant Seebeck value and a high ZT value of 2.4 has been attained in the case of 2D electron gas in Sr$_2$TiO$_3$ \cite{perovskites_8}.

In the meantime, one more emerging class of 2D metal oxides have gained much attention due to their potential applications \cite{metal_oxides_1}. Their eco-friendly nature with low toxicity, good photothermal response and attractive electrochemical properties lead them to be efficient energy conversion and storage materials \cite{metal_oxides_2, metal_oxides_3, metal_oxides_4, metal_oxides_5, metal_oxides_6}.

\section{Comparison of the ZT between the 2D and other materials}
\begin{figure*}
    \centering   
    \includegraphics[width=1.0\textwidth]{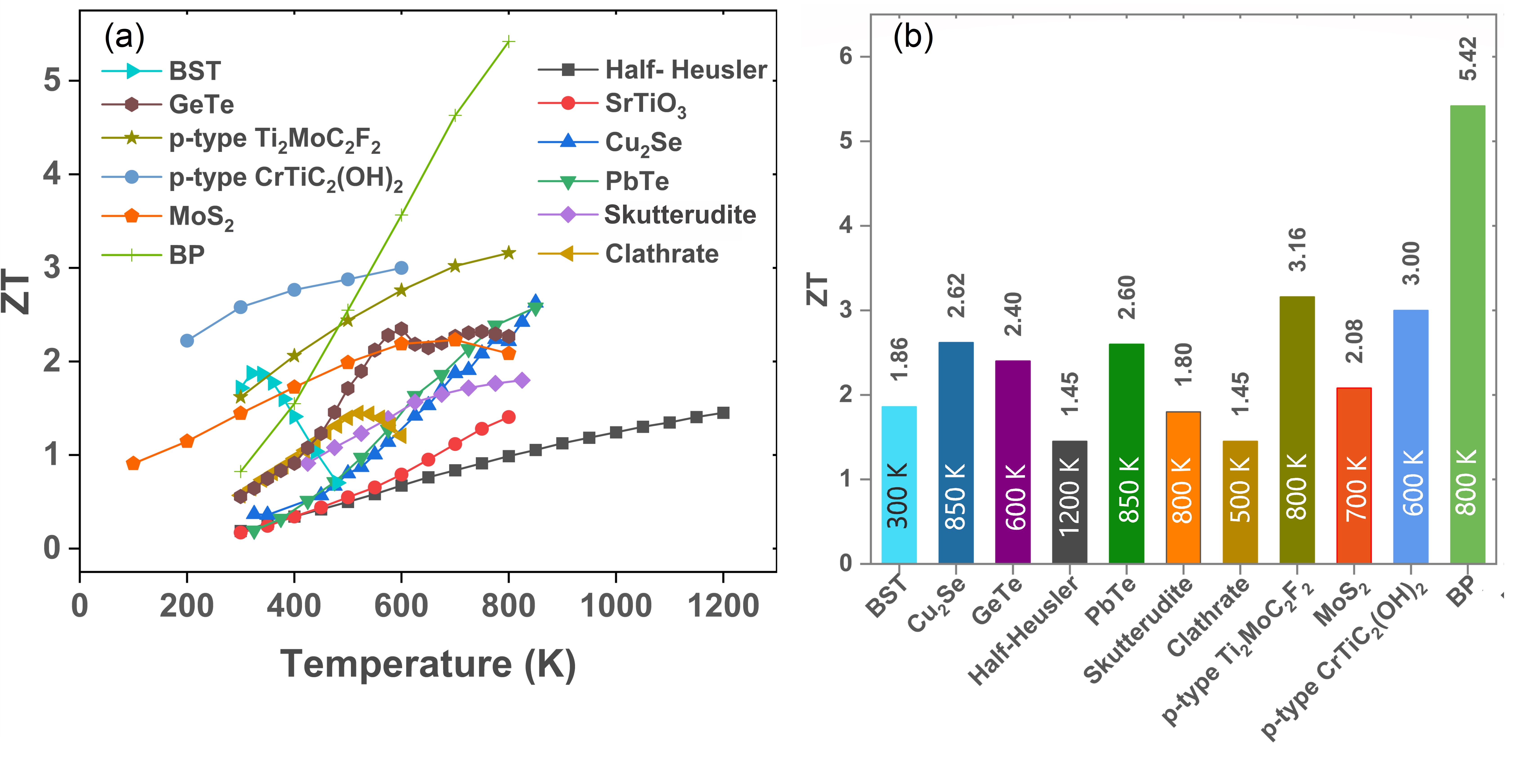}
    \caption{(a) ZT values of well established TE materials BST \cite{BST_1}, GeTe \cite{GeTe}, MoS$_2$ \cite{MoS2_nanoribbons}, BP \cite{BP_TE_ZT_1}, half-Heusler \cite{half-Heusler_fu_realizing_2015}, SrTiO$_3$ \cite{SrTiO3_WANG2017387}, Cu$_2$Se \cite{Cu2Se_Olvera_2017}, PbTe \cite{PbTe_WU20191276}, skutterudite \cite{skutterides_Rogl2015}, clathrate \cite{clathrates_SAIGA2012303} and comparison to few potential MXenes, p-type Ti$_2$MoC$_2$F$_2$ \cite{Shiladitya-Karmakar_TE_2020} and Cr$_2$TiC$_2$(OH)$_2$ \cite{Ziang-Jing_TE_2019} as a function of temperature, (b) Highest values of ZT attained by different TE materials and MXenes}
    \label{Figure11}
\end{figure*}

\begin{figure}
    \centering
    \includegraphics[width=\columnwidth]{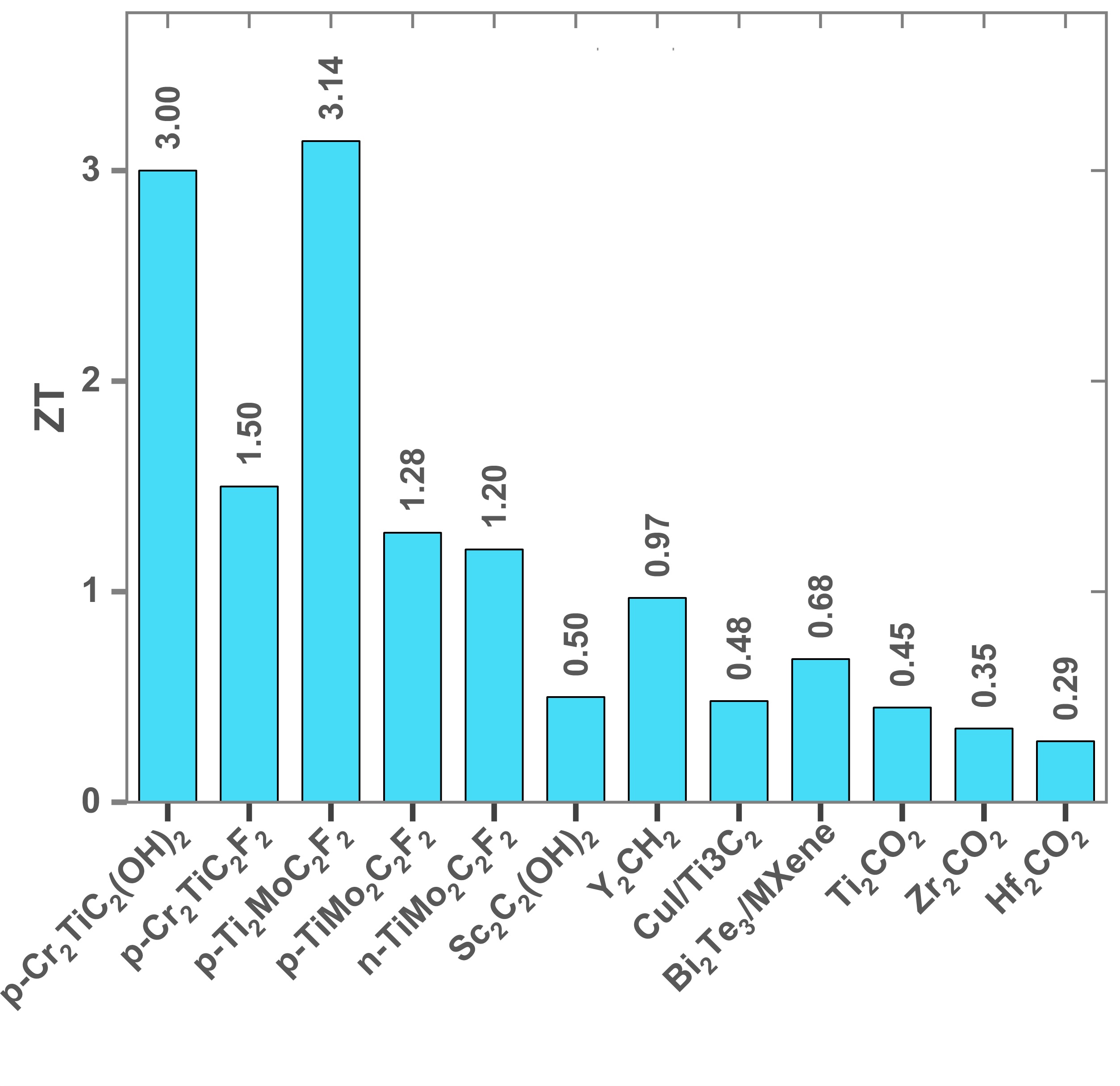}
    \caption{Maximum ZT values of different MXenes at different temperatures reported in this review}
    \label{MXenes_ZT}
\end{figure}

The theoretical predictions of the semiconducting MXenes revealed that the TE properties are comparable to that of the conventional TE materials as shown in Figure \ref{Figure11}a. The highest values of ZT attained by MXenes Ti$_2$MoC$_2$F$_2$ \cite{Shiladitya-Karmakar_TE_2020} and Cr$_2$TiC$_2$(OH)$_2$ \cite{Ziang-Jing_TE_2019} have been compared with the existing conventional bulk and 2D TE materials like Bi$_{0.5}$Sb$_{1.5}$Te$_3$ (BST) \cite{BST_1}, GeTe (Ge$_{0.86}$Pb$_{0.1}$Bi$_{0.04}$Te) \cite{GeTe}, MoS$_2$ \cite{MoS2_nanoribbons}, BP \cite{BP_TE_ZT_1}, half-Heusler (FeNb$_{0.88}$Hf$_{0.12}$Sb) \cite{half-Heusler_fu_realizing_2015}, SrTiO$_3$ \cite{SrTiO3_WANG2017387}, Cu$_2$Se \cite{Cu2Se_Olvera_2017}, PbTe (Na$_{0.03}$Eu$_{0.03}$Sn$_{0.02}$Pb$_{0.92}$Te) \cite{PbTe_WU20191276}, skutterudites \cite{skutterides_Rogl2015}, and Clathrate (Cu-doped Ba$_8$Ga$_{16}$Sn$_{30}$) \cite{clathrates_SAIGA2012303} respectively. A high ZT value of around 7.38 has been predicted for the n-type half-Heusler compound BCaGa at 700 K \cite{half-Heusler_highZT}. The maximum ZT values by these materials at different temperatures have been also compared and displayed in Figure \ref{Figure11}b. A maximum ZT value of around 3 has been obtained with the MXenes. Figure \ref{MXenes_ZT} shows the maximum ZT values attained by different MXenes at different temperatures. The different TE parameters attained by several MXenes discussed in this review have been listed in Table \ref{tab:MXene_TE}.

\section{Extension of TE application to MBenes: possibilities and prospects}

\begin{table*}[hbt!]
\caption{Thermoelectric properties of several MXenes } \label{tab:MXene_TE}
 \def\arraystretch{1.5}
\begin{center}
\begin{tabular}{|c|c|c|c|c|c|c|c|}
  \hline
 & T & S & $\sigma$  & PF  & $\kappa$  & ZT & Ref\\ 
 &(K)& ($\mu$VK$^{-1}$)&(Sm$^{-1}$)&(Wm$^{-1}K^{-2}$) & (Wm$^{-1}$K$^{-1}$) & & \\ \hline
 Ti$_2$CO$_2$ & 100 & 1140 &  &  &  &  & \cite{https:Khazaei_TE_2013}\\ \hline
 Sc$_2$C(OH)$_2$ & 100 & $\approx$ 2000 & & & & & \cite{https:Khazaei_TE_2013}\\ \hline
 Ti$_2$CO$_2$ & 300 &  &  &  & 21.9 &  & \cite{Gandi_TE_2016}\\ \hline
 Zr$_2$CO$_2$ & 300 &  &  &  & 61.9 &  & \cite{Gandi_TE_2016}\\ \hline
 Hf$_2$CO$_2$ & 300 &  &  &  & 86.3 &  & \cite{Gandi_TE_2016}\\ \hline
 Sc$_2$C(OH)$_2$ & 300 & 372 &  &  & 10 &  & \cite{Kumar_TE_2016}\\ \hline
 Sc$_2$CO$_2$ & 300 & 1022 &  &  & 59 &  & \cite{Kumar_TE_2016}\\ \hline
 Sc$_2$CF$_2$ & 300 & 1036 &  &  & 36 &  & \cite{Kumar_TE_2016}\\ \hline
 Mo$_2$Ti$_2$C$_3$T$_2$ & 803 & -47.3 & 13.8 $\times$ 10$^{4}$ & 3.09 $\times$ 10$^{-4}$   &  &  & \cite{Kim_TE_2017}\\ \hline
 Ti$_2$CO$_2$ & 300 &  &  &  & 43.95 &  & \cite{Zhonglu_TE_2018}\\ \hline
 Ti$_3$C$_2$T$_x$ film & 300 & 16.5 & 1.652 $\times$ 10$^5$ & 44.98 $\times$ 10$^{-6}$ &  &  & \cite{Peng-Liu_TE_2020}\\ \hline
 Mo$_2$CF$_2$ & 100 & 995 &  &  &  &  & \cite{Xiangda-Zou_TE_2018}\\ \hline
 Cr$_2$CF$_2$ & 100 & 1046 &  &  &  &  & \cite{Xiangda-Zou_TE_2018}\\ \hline
 Y$_2$CH$_2$ & 100 & 1188 &  &  &  & 0.97 & \cite{OMUGBE_TE_2022}\\ \hline
 n-Cr$_2$TiC$_2$ & 300  & $\approx$ 930 & 9.530 $\times$ 10$^5$ &  & 150 & 0.4 & \cite{Ziang-Jing_TE_2019}\\ \hline
 p-Cr$_2$TiC$_2$F$_2$ & 300 & $\approx$ 450 & 3.118 $\times$ 10$^5$ &  & 39 & 1.0 & \cite{Ziang-Jing_TE_2019}\\ \hline
 p-Cr$_2$TiC$_2$(OH)$_2$ & 300  & $\approx$ 500 & 2.127 $\times$ 10$^6$ &  & 6.5 & 2.59 & \cite{Ziang-Jing_TE_2019}\\ \hline
 p-Ti$_2$MoC$_2$F$_2$ & 800 & 150 & $\approx$ 1.2 $\times$ 10$^5$ &  & 0.3 & 3.1 & \cite{Shiladitya-Karmakar_TE_2020}\\ \hline
 n-TiMoCO$_2$ & 300  &  &  & 4.98 $\times$ 10$^{-7}$ &  &  & \cite{Zicong-Wong_TE_2020}\\ \hline
p-TiMoCO$_2$ & 300  &  &  & 1.5 $\times$ 10$^{-7}$ &  &  & \cite{Zicong-Wong_TE_2020}\\ \hline
 Ti$_3$C$_2$ flakes &  &  &  &  &  & 0.112 & \cite{Y.I.Jhon_TE_2018}\\ \hline
 Ti$_3$C$_2$T$_x$-SWCNT & 300 & -32.2 & 7.509 $\times$ 10$^4$  & 77.9 $\times$ 10$^{-6}$ &  &  & \cite{Wenjun-Ding_TE_2020}\\ \hline
 Bi$_2$Te$_3$/MXene & 380 &  &  & 1.49 $\times$ 10$^{-3}$ & 0.41 & 0.68 & \cite{Dewei-ZHANG_TE_2022}\\ \hline
 Ti$_3$C$_2$T$_x$/BST composite & 300 & 192 & 6.8 $\times$ 10$^4$ & 25 $\times$ 10$^{-4}$ & 0.65 & 1.15 & \cite{BST-Ti3C2Tx}\\ \hline
(Sc$_{2/3}$Cd$_{1/3}$)C & 300 & 820 &  &  &  &  & \cite{Qiang-Gao_TE_2020}\\ \hline
(Sc$_{2/3}$Hg$_{1/3}$)C & 300 & 1200 &  &  &  &  & \cite{Qiang-Gao_TE_2020}\\ \hline
MXene/Polymer & 300 & 57.3 &  & 155 $\times$ 10$^{-6}$ &  &  & \cite{Xin-Guan_TE_2020}\\ \hline
CuI/Ti$_3$C$_2$ & 550 &  &  & 225 $\times$ 10$^{-6}$ &  & 0.48 & \cite{Vaithinathan-Karthikeyan_TE_2022}\\ \hline
p-Mo$_2$C thin film & 300 & 308 & 0.22 &  & 0.37 &  & \cite{Taeho-Park_TE_2021}\\ \hline
n-Mo$_2$Ti$_2$C$_3$ thin film & 300 & -25 & 628 &  & 0.45 &  & \cite{Taeho-Park_TE_2021}\\ \hline
\end{tabular}
\end{center}
\end{table*}

The quest is still ongoing for a lot of several other materials for different applications. Recently, a novel kind of 2D materials, MBenes which are similar to MXenes in many ways have been discovered \cite{MBenes_discovery, MBenes_6}. The 2D nanosheets of early transition metal borides are labeled as MBenes and can be derived from the MAB phases where B is Boron which is shown in Figure \ref{PeriodicTable}. Similar to MXenes, MBenes can be prepared by etching the surface of the MAB phase. This led to the formation of many new phases that crystallize into orthorhombic, hexagonal, and tetragonal MBenes with different stoichiometric ratios of metal and boron atoms \cite{MBenes_Varun}. This new category of 2D materials grabbed the attention of scientists to analyze them experimentally as well as theoretically in different areas of science. A wide range of research has been accomplished in the past few years to explore the structural, mechanical, electronic, magnetic properties and the applications of MBenes \cite{https:Khazaei_TE_2013, Khaledialidusti_2021, MBenes_1, MBenes_2, MBenes_Varun, MBenes_4, MBenes_5, MBenes_Agnieszka}.

Even though MBenes are similar to MXenes, they vary in many ways due to distinct stoichiometries, different orientations of 2D layers, and variable types of structural transformations \cite{MBenes_structures, MBenes_Varun}. All pristine MXenes possess the hexagonal crystal structure and belong to the \textit{P6$_3$/mmc} space group, whereas  MBenes can exhibit additional crystal symmetries such as orthorhombic and tetragonal ones. In particular, within the orthorhombic symmetry the in-plane lattice parameters a and b are not equivalent, leading to in-plane direction-dependent properties as widely discussed in Sec. \ref{sec:anisotropy} for other 2D systems. In contrast to MXenes, an in-plane structural anisotropy is an additional factor in likely enhancing the TE properties of particular MBenes structures. 

The advancement of theories and computational methods has been remarkably beneficial in analyzing TE properties. Indeed in the last two decades, the accuracy and productivity of density functional theory (DFT) based calculations have improved to a great extent. As it is required to work on several MBenes, the adaption of high throughput computation and Machine learning techniques can be utilized \cite{ML_1, ML_2, ML_3, ML_4, ML_5}. In particular, Machine learning is grabbing all the attention as it is emerging in the computational field in understanding and predicting many new materials. In recent years, machine learning techniques have been applied successfully applied on the TE materials \cite{ML_TE_1, ML_TE_2, ML_TE_3, ML_TE_4, ML_TE_5}. However, it is necessary to flourish machine learning methods for fruitful endeavors in the TE field.

Even though few MBenes are theoretically predicted, hardly a very small number of MBenes have been synthesized so far. The research performed on the TE properties of MBenes to date is very limited. In this context, there is a lot of scope to investigate MBenes for energy storage applications including thermoelectricity. It is aimed to theoretically analyze several MBenes to predict their TE performance with band engineering and functionalization and to implement several other techniques to unite them with highly efficient TE materials. Few theoretical works have already started investigating the TE efficiency in MBenes.

The thermal transport of Mo-based MBene has been analyzed by Yipeng An \textit{et al} for the first time \cite{Yipeng-An_TE_2019}. The electronic structure calculations show that the MoB$_2$ monolayer exhibits metallic nature. The lattice thermal conductivity of MoB$_2$ monolayer of around 20 angstroms has been estimated to be 11.54 $Wm^{-1}K^{-1}$ at room temperature by solving harmonic and anharmonic interatomic force constants. The ultralow lattice thermal conductivity value indicates that this monolayer could be utilized as a heat insulating material.

Zha \textit{et al} examined the electrical and thermal conductivities of MBene Mo$_2$B in both H-type and T-type configurations for lithium ion batteries with the aid of first principle calculations \cite{Xian-Hu-Zha_TE_2020}. The electrical conductivities for both the structures of Mo$_2$B have been found to be in the order of $10^6$ $\ohm^{-1}m^{-1}$. By considering a flake length of around 5 $\mu$m, the thermal conductivities along the armchair direction of the H-type and T-type structures have been computed as 197.9 and 158.3 $Wm^{-1}K^{-1}$ respectively at room temperature. These high values of thermal conductivities indicate the importance of engineering these compounds to utilize them as TE compounds.

\subsection{Possibilities to enhance TE efficiency in MBenes}

This section will discuss the possible ways to enhance the ZT value of MBenes for future works. The magnitude of ZT can be improved by two different approaches. The electronic structure can be modified to optimize the Seebeck coefficient, electrical conductivity, and electronic thermal conductivity. The lattice thermal conductivity can be reduced by engineering the phonon path. It is well known that semiconductors with narrow band gaps or pseudo-gap systems are the best TE materials at room temperature \cite{Sofo_Optimumbandgap}.

It is very clearly evident from the discussion so far that the electronic structure of MXenes has been altered from metallic to semiconducting with functionalization. The same can be applied to MBenes and their TE nature can be investigated. However, it has to be noted that the functionalization of Cr$_2$B$_2$ MBenes with -O, -OH, -Cl, or -H exhibited metallic nature \cite{Cr2B2T2_Hu_functionalization}. Meanwhile, the electronic structure calculations performed by Khaledialidusti \textit{et al} show that Sc$_2$B$_2$O$_2$ to be a semiconductor, while, Sc$_2$B$_2$F$_2$, Sc$_2$B$_2$(OH)$_2$, Zr$_2$B$_2$O$_2$ and Hf$_2$B$_2$O$_2$ turned out to be semimetallic \cite{Khaledialidusti_2021}. Unlike MXenes, very few compounds turned out to be semimetallic and very few turned out to be semiconducting with functionalization. Nevertheless, it must be emphasized that the literature review suggests that semimetals can also be good candidates for TE applications due to their high Seebeck coefficient values. Concerning the prominence of the electronic band structure's impact on the TE properties, it is noteworthy to mention the "pudding-mold" type found in highly efficient TE materials such as Na$_x$CoO$_2$ \cite{pudding_Kuroki} and CuAlO$_2$ \cite{pudding_Mori}. The high TE power factor value of the synthesized MXene, Mo$_2$TiC$_2$T$_x$ has been also associated with the pudding-mold type band structure \cite{Kim_TE_2017}. Thus it is important to explore this type of band phenomenon in MBenes too.

It must also emphasize that the bulk (3D) transition metal borides are well known for their high temperature TE applications \cite{3D_TMD_3}. The metallic nature of the 3D transition metal borides can be turned to semiconducting by increasing the boron content \cite{3D_TMB_1, 3D_TMB_2}. So, in a similar fashion, semiconducting MBenes can also be obtained. However, the possibility of boron-rich MBenes has yet to be examined.
\begin{figure*}
    \centering
\includegraphics[width=0.75\textwidth]{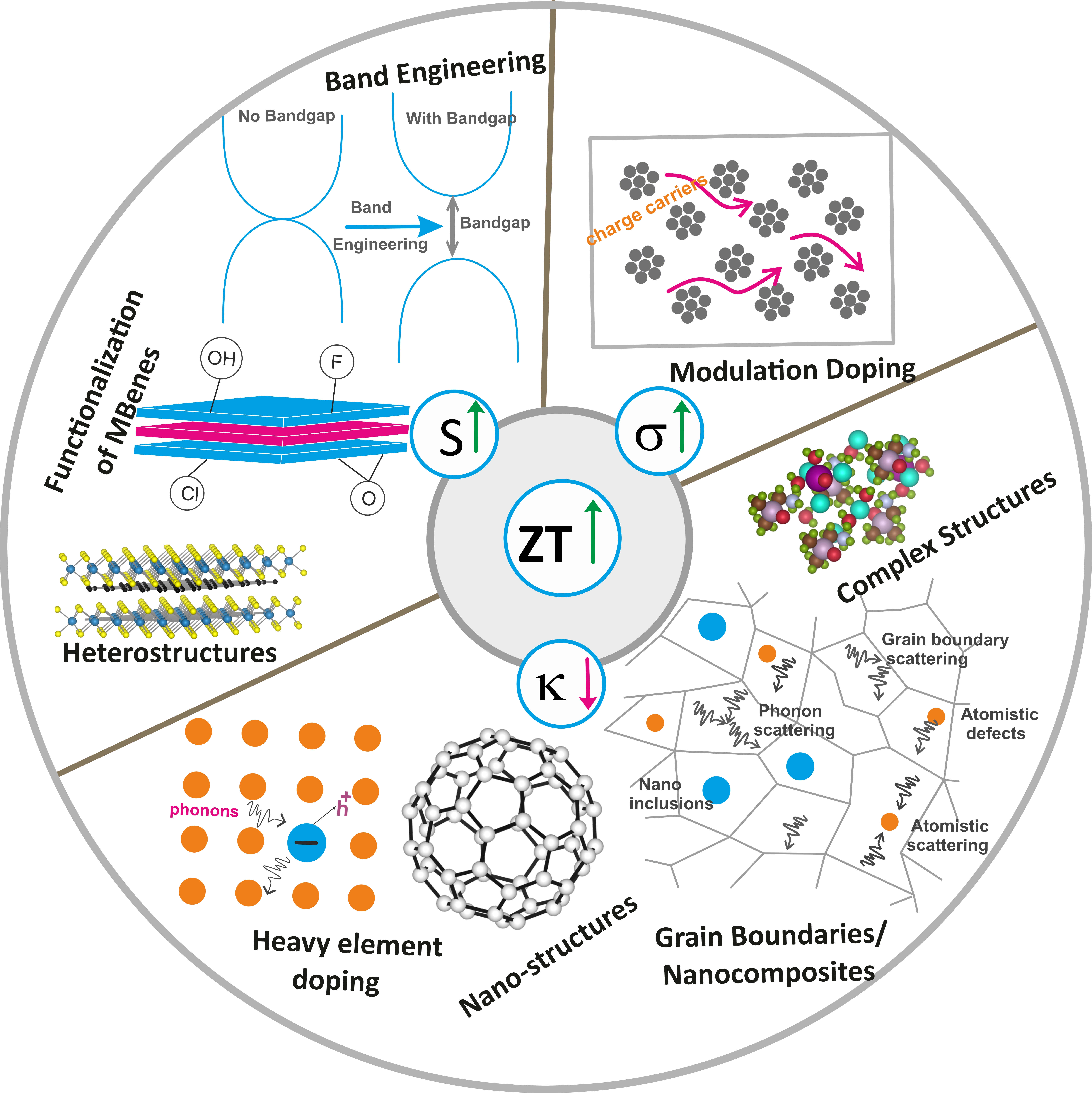}
    \caption{Different approaches to enhance ZT values of MBenes}
    \label{ZT-prospects}
\end{figure*}

Now it is well established that the band structure of the compounds shows a great impact on the TE nature, the band gap of the MBenes can be also modified by band engineering. The band structure of MBenes can be engineered among others by alloying or resonant doping. Band engineering changes the effective mass of carriers and can optimize carrier concentration and increase carrier mobility resulting in greater power factor values. However, aimlessly increasing the effective mass can reduce carrier mobility and unwillingly impact the ZT values. Therefore, band engineering has to be implemented in such a way as to attain a balance between carrier mobility and effective mass to yield positive results.

One more concept of enhancing the Seebeck coefficient is by tailoring the stoichiometry as well as the microstructure. The most probable defects possible in the crystal structures of MBenes have to be identified with the aid of electronic structure calculations. The off-stoichiometric compounds have to be investigated to find out their influence on the TE nature.  In both the cases of MXenes as well as MBenes, different crystallographic orientations are possible.

One of the most conventional ways used in the optimization of TE materials is doping. Doping always tends to optimize the carrier concentration which can improve the power factor values\cite{Zebarjadi_TEperspectives}. The doping can be applied in three different approaches: modulation doping, uniform doping, and gradient doping. Modulation doping is mostly implemented in the field of microelectronics and photonic devices which can also be applied to the MBenes. Instead of doping all over the sample uniformly, only one type of nanograins will be doped with modulation doping. This will result in the spatial separation of undoped and nanograins and lead to the reduction of electron scattering and higher mobility. As a result of enhancement in the carrier mobility, the electrical conductivity can be increased. While, in the case of uniform or conventional doping, the change in temperature does not influence the carrier concentration which doesn't assure the TE performance. This can be overcome by temperature-dependent gradient doping.

Another factor that can influence TE efficiency is heterostructuring. It must be emphasized that the power factor values in MXenes have been enhanced by the formation of heterostructures. The electronic structure of the MBenes can also be modified with this method. However, the weak vdW bonding between the layers in the stacking 2D heterostructures can lead to the reduction of electrical conductivity. 

Meanwhile, the mean free path or the relaxation time of the phonons must be as small as possible to knock off the thermal conductivity that can be achieved with a perturbation in the periodicity of the crystal lattice. The compact crystal structure approach consisting of heavy elements can be utilized to attain this. In the case of solid solutions or point defects or dislocations or even nanoparticles, the thermal conductivity can be modified with the addition of phonon diffusion centers like atoms with large atomic masses. Different aspects that can be followed to reduce the thermal conductivity have been shown in Figure \ref{ZT-prospects}. Nanostructuring of MBenes can lead to the diffusion of phonons which will not only reduce the thermal conductivity but can also enhance the Seebeck coefficient due to the modification of DOS by the asymmetry associated with the quantum confinement of the carriers. It is evident from the review of MXenes that the nanoinclusions or nanocomposites can impact the overall TE performance by reducing the thermal conductivity. The same can be applied to MBenes as well. 

One more promising way of reducing the thermal conductivity of MBenes is by heavy element doping. Heavy element doping can modify the carrier concentration by launching intense point defect scattering of phonons and resulting in a reduction of lattice thermal conductivity. Complex crystal structures can also contribute to the same cause by significantly reducing the range of acoustic phonons that impacts the values of lattice thermal conductivity. There might also be several other ways that can be implemented to enhance the TE behavior of MBenes. There is a lot of scope to perform a great extent of research on these materials and to explore the TE properties.

Recently, a review article based on interface/surface modification illustrates how the discontinuous interfacial modification in the MXenes can improve the transport properties \cite{Shiyang-He_TE_2021}. The thermal conductivity can be reduced with high interface densities attained from nano- or micro-scale grain boundaries. The power factor of TE materials can be enhanced by modifying the electronic structure. This review article discussed in detail how the TE performance of ZnO and Bi$_{0.4}$Sb$_{1.6}$Te$_3$ can be enhanced by Ti$_3$C$_2$T$_x$ MXene dispersion \cite{GuO_interface_TE_MXene}. Likewise, the capability of MBenes can also be verified if they can be utilized as nanocomposites in interface/surface modification and boost the TE performance.

Along with the several aspects that can show influence on the TE properties, it is also important to consider a few important factors. The TE properties greatly depend on the layer thickness of the 2D materials as it is evident from the case of TMDs \cite{TMDC_TE_3}. This shows the significance of the layer thickness of MBenes while evaluating their TE properties.

\section{Challenges, gaps  in current research and  limitations in TE field}
Here, we listed some crucial aspects of TE research, starting from the limitations in the theoretical description, and ending with the essential gap in TE research regarding MXenes materials in the context of newly discover MBenes structures.

Regardless, of the successful endeavors of TE modeling, it is essential to discuss the challenges and limitations of these methods in evaluating the TE entities. The nature and accuracy of the properties of TE materials obtained from the first principle calculations are quite essential. Coming to DFT methods, it is well known that the estimation of TE coefficients depends upon solving the Boltzmann transport equation (BTE) under the assumption of relaxation time ($\tau$) approximation \cite{Sofo, BoltzTraP}. The $\tau$ that is based on the scattering mechanism and temperature dependent \cite{tau_temp_dependent_1, tau_temp_dependent_2}, is generally assumed to be constant in this computational approach. Even though the Seebeck values do not depend upon the $\tau$, one of the limitations of this theoretical methodology is that the electrical and electronic thermal conductivities can only be obtained either as a function of $\tau$ or by fitting to the experimentally existing values \cite{KKR-CPA}. Assuming $\tau$ as constant can also result in big differences between the theoretical and experimental values. It is quite important to further develop methods that can easily determine $\tau$ and overcome this limitation for better accuracy of the results.

The transport properties can be significantly affected in the case of n- or p-doped materials within the simple and commonly used rigid band approximation (RBA) \cite{PhysRevB.74.155205}. Within this approach, one needs to assume the electronic structure of the doped system is not changed (only the chemical potential changes with the doping concentration and temperature). However, the dopants usually modify the electronic density of states (DOS), hence, affecting the conductivities \cite{Fang2017}. Another aspect regarding the doped materials is a  supercell approach implemented in most of the  DFT software. Namely, the supercell reduces the crystal symmetry of the doped systems,  which is usually not observed in experiments such as  X-ray or electron diffraction. While real systems neither possess ideal symmetry nor transform into perfect supercells, the impact of the symmetry on the band structure is also important. This issue can be overwhelmed for instance by using the Korringa-Kohn-Rostoker Green function formalism under the coherent potential approximation (KKR-CPA).

In recent years, the data-driven and machine learning methods procure the utmost importance in materials science due to their significance in scrutinizing enormous amounts of material data. In this context, data-driven methods have been also applied in the field of thermoelectricity as a part of discovering and predicting a wide range of promising TE materials \cite{data_driven_methods_TE}. However, the main challenge in these data-driven methods, especially for machine learning, is that it requires a large set of databases obtained from high-throughput computations and/or experimentally determined parameters for accurate predictions. Even though the existing high-throughput methods are very effective in describing the electron and phonon transport mechanisms for the simple TE compounds, but fail in dealing the complex structural systems and at high temperature ranges. This leads to the necessity of encouraging machine learning tools to evolve more into a greater extent in data acquisition.   

As a matter of prospects, a same similar path can be paved to the last addition to the flatland of 2D materials: 
 MBenes. The priorities that should be taken into account in selecting the materials for thermoelectric applications and different ways to optimize the efficiency of MBenes have been reviewed in detail. 
Furthermore, the magnetic nature of a few MBenes cannot be neglected as the transport mechanism not only by the electrons but also via magnetic spins is also essential. The well know definition of the Seebeck effect is the generation of voltage with the temperature gradient between two conductors. Likewise, the phenomenon of production of the spin voltage as a result of spin-polarized conduction electrons caused by the difference in the temperature of the ferromagnet is known as the spin Seebeck effect \cite{SpinSeebeckeffect_1}. In recent times, the spin transport mechanism with the combination of electronic conduction is highly anticipated in the field of thermoelectrics and a crucial research criterion for the spintronic applications of the 2D vdW magnetic materials \cite{SpinSeebeckeffect_2}. In this scenario, it is noteworthy to bring up that the spin Seebeck effect has not been considered in the study of TE properties of magnetic MXenes. Thus, it is also important to consider the spin Seebeck effect also while dealing with the TE nature of magnetic MBenes.

Meanwhile, in the last two decades, several synthesis strategies have been implemented successfully to fabricate high-performance TE materials with high ZT values. Nonetheless, the scalability factor remains to be restricted as the production of TE materials from the lab scale to the industrial scale is highly limited due to various factors. To overcome this limitation and to compete with other technologies, the TE research has to be directed at manufacturing TE devices for large scale industrial production. On the other hand, there will be several challenges in the coming years to synthesize MXenes and MBenes. At the moment, the majority of the syntheses of MXenes and MBenes rely on their parent precursor MAX and MAB phases respectively. It is essential to limit the use of MAX and MAB phases as they consume larger amounts of energy for their synthesis and can greatly affect the environment. Moving ahead, there is a need for better elemental composition, a safer manufacturing process, reduced costs, and increased mass production.

The existing thermal property measurements can lead to the electromigration or diffusion at the interconnecting or conducting stripes of the MXenes or MBenes. This will deteriorate the TE performance of the materials in long-term operations. Therefore, more appropriate techniques are needed to improve the thermal property measurements at the working temperature. To convert the human body heat into electrical power even more efficiently, it becomes imperative to focus on the development of lightweight, conformable, and breathable MXene and MBene-based composites to fulfill the practical demands of wearable TE devices. The designing of TE devices must sincerely consider the multi-field effect. This is because the heat sources induced from the photo, magnetic, or microwave sources can introduce additional factors that may influence the performance of the TE device and may be used for space applications in the future where the presence of a gravity field cannot be assumed.

The realization of MXenes' and MBenes' full potential as TE materials require the implementation of more comprehensive theoretical and experimental approaches. The integration of both theoretical and experimental investigations is expected to significantly promote the TE performance of MXenes and MBenes. The diverse demands of clean energy conversion or storage and the development of flexible electronic devices can be accomplished.

\section{Conclusion}

 In summary, different aspects and the importance of thermoelectrics have been briefly discussed. In the search for novel and sustainable forms of energy, thermoelectricity gained the limelight of several researchers as they believed it to be one of the leading technologies in the future. Thermoelectrics can assist in recovering energy from waste heat as well as generate power. However, it is necessary to design new materials with higher efficiencies than the existing ones. It is quite challenging to increase the figure of merit ZT of TE materials due to a competing relation between the parameters involved in it. 

The main motto of this review article is to present the accomplishments of MXenes in the field of thermoelectrics. The TE properties of various MXenes existed in the literature have been explained in a very detailed manner. The TE performance of MXenes has been enhanced with the aid of different approaches like band engineering, heterostructuring, and other nanostructuring techniques such as nano-inclusions and nanocomposites. This strategy simply leads to the enlargement of the power factor by tuning the electronic structure or engineering the phonon paths by introducing disorder (defects, vacancies, etc.). The latter increases the phonon scattering and overall reduces their thermal conductivity. In addition, other aspects such as layer-dependent TE properties and structural anisotropy, which lead to the orientation dependent TE behavior, pave the way in device design for potential TE applications. The commercial applications of TE devices have been also partially covered, with possibilities and prospects extended to MBenes.
 
\section*{Author Contributions}
S.B.  collected and analyzed the data; M.B. developed the concept of the study, designed the structure of the manuscript, supervised the research as a project leader, and coordinated the preparation of the manuscript. All authors helped to prepare, correct, and proofread the manuscript.

\section*{Acknowledgements}
We acknowledge support from the  National Science Centre (NCN Poland) within grant UMO-
2019/35/B/ST5/02538. M.B. acknowledges support from the University of Warsaw within the project “Excellence Initiative-Research University” program. Access to computing facilities of PL-Grid Polish Infrastructure for Supporting Computational Science in the European Research Space, the Wroclaw Centre for Networking, and the Interdisciplinary Center of Modeling (ICM), University of Warsaw are gratefully acknowledged. We acknowledge ACK Cyfronet AGH (Poland) for awarding this project access to the LUMI supercomputer, owned by the EuroHPC Joint Undertaking, hosted by CSC (Finland) and the LUMI consortium through Pl-Grid organization (Poland), under the grant entitled: "Electronic, optical and thermoelectric properties of selected layered materials and selected heterostructures". We would like to thank Mr. Varun Gopalakrishnan Nair from Warsaw University of Technology, Faculty of Materials Science and Engineering for his assistance in creating some figures for this review. 

\bibliography{main}

\end{document}